\documentclass[sigconf]{acmart}

\copyrightyear{2024}
\acmYear{2024}
\setcopyright{acmlicensed}
\acmConference[CCS '24] {Proceedings of the 2024 ACM SIGSAC Conference on Computer and Communications Security}{October 14--18, 2024}{Salt Lake City, UT, USA.}
\acmBooktitle{Proceedings of the 2024 ACM SIGSAC Conference on Computer and Communications Security (CCS '24), October 14--18, 2024, Salt Lake City, UT, USA}
\acmISBN{979-8-4007-0636-3/24/10}
\acmDOI{10.1145/3658644.3690269}

\newtheorem{mydef}{Def}
\usepackage{subfigure}
\usepackage{graphicx}
\usepackage{enumitem}
\usepackage{pifont}
\usepackage{multirow,makecell}
\usepackage{tcolorbox}
\usepackage{color,xcolor}
\usepackage{listings,amsfonts}
\usepackage{caption}
\usepackage{subcaption}
\usepackage{threeparttable}
\usepackage{bbding}
\usepackage{booktabs} 
\usepackage{longtable}
\usepackage{flushend}
\usepackage{xspace}
\usepackage{wasysym}
\usepackage[ruled,vlined,linesnumbered]{algorithm2e}
\usepackage{amsmath}
\usepackage{balance}

\newcommand{\toolname}{\textsc{CanCal}\xspace}

\newcommand{\crappendix}[1]{\textcolor{purple}{Appendix~#1}}

\AtEndPreamble{
	\usepackage{hyperref}

	\hypersetup{
		colorlinks = true,
		linkcolor = purple,
		anchorcolor = purple,
		citecolor = purple,
		filecolor = purple,
		urlcolor = purple
	}
}

\begin{document}

\title{\toolname: Towards Real-time and Lightweight Ransomware Detection and Response in Industrial Environments}

\author[S Wang]{Shenao Wang}
\authornote{Both authors contributed equally to this research.}
\authornote{Hubei Key Laboratory of Distributed System Security, Hubei
Engineering Research Center on Big Data Security, School of Cyber Science and
Engineering, Huazhong University of Science and Technology.}
\email{shenaowang@hust.edu.cn}
\affiliation{%
  \institution{Huazhong University of Science and Technology}
  \city{Wuhan}           
  \country{China}
}

\author[F Dong]{Feng Dong}
\authornotemark[1]
\authornotemark[2]
\email{dongfeng@hust.edu.cn}
\affiliation{%
  \institution{Huazhong University of Science and Technology}
  \city{Wuhan}           
  \country{China}
}

\author[H Yang]{Hangfeng Yang}
\email{yanghangfeng@sangfor.com.cn}
\affiliation{%
  \institution{Sangfor Technologies Inc.}
  \city{Shenzhen}           
  \country{China}
}

\author[J Xu]{Jingheng Xu}
\authornote{Co-corresponding authors.}
\email{xjh@sangfor.com.cn}
\affiliation{%
  \institution{Sangfor Technologies Inc.}
  \city{Shenzhen}           
  \country{China}
}

\author[H Wang]{Haoyu Wang}
\authornotemark[2]
\authornotemark[3]
\email{haoyuwang@hust.edu.cn}
\affiliation{%
  \institution{Huazhong University of Science and Technology}
  \city{Wuhan}           
  \country{China}
}

\begin{abstract}
Ransomware attacks have emerged as one of the most significant cybersecurity threats. Despite numerous methods proposed for detecting and defending against ransomware, existing approaches face two fundamental limitations in large-scale industrial applications: (1) Behavior-based detection engines suffer from the enormous overhead of monitoring all processes and resource constraints for model inference, failing to meet the requirements for real-time detection; (2) Decoy-based detection engines generate an overwhelming number of false positives in large-scale industrial clusters, leading to intolerable disruptions to critical processes and excessive inspection efforts from security analysts. To address these challenges, we propose \toolname, a real-time and lightweight ransomware detection system. Specifically, instead of indiscriminately analyzing all processes, \toolname selectively filters suspicious processes by the monitoring layers and then performs in-depth behavioral analysis to isolate ransomware activities from benign operations, minimizing alert fatigue while ensuring lightweight computational and storage overhead. The experimental results on a large-scale industrial environment~(1,761 ransomware, \string~ 3 million events, continuous test over 5 months) indicate that \toolname achieves a remarkable 99.65\% true positive rate on 555,678 unknown ransomware behavior events, with near-zero false positives. \toolname is as effective as state-of-the-art techniques while enabling rapid inference within 30ms and real-time response within a maximum of 3 seconds. \toolname dramatically reduces average CPU utilization by 91.04\% (from 6.7\% to 0.6\%) and peak CPU utilization by 76.69\% (from 26.6\% to 6.2\%), while avoiding 76.50\% (from 3,192 to 750) of the inspection efforts from security analysts. By the time of this writing, \toolname has been integrated into a commercial product and successfully deployed on 3.32 million endpoints for over a year. From March 2023 to April 2024, \toolname successfully detected and thwarted 61 ransomware attacks. A detailed manual forensic analysis of 27 ransomware attacks from March to June 2023~(including 13 n-day exploits and 5 high-risk zero-day attacks) demonstrates the effectiveness of \toolname in combating sophisticated and unknown ransomware threats in real-world scenarios.
\end{abstract}



\maketitle

\section{Introduction}
Recently, ransomware attacks have become one of the biggest threats in the field of network security.
With the rise of Ransomware as a Service~(RaaS), cybercriminals can launch attacks against individual users, enterprises and governments~\cite{sophos2023ransomware,chainalysiss2024ransomware}. These attackers usually adopt sophisticated tactics and techniques~(e.g., 0-day exploitation~\cite{steve20230day}, file-less attacks~\cite{kurt2023fileless}, and process injection~\cite{mosimilolu2020injection}), making the ransomware more targeted and covert~\cite{DBLP:journals/access/RazaullaFMGMFA23}. 
Traditional signature-based Endpoint Detection and Response~(EDR) systems~\cite{DBLP:conf/dasc/MedhatGA18, DBLP:conf/icacci/SheenY18, DBLP:journals/compsec/ZhuJSWAC22} rely on identifying known patterns within binary files, which is prone to be evaded by adversarial techniques, such as obfuscation and polymorphism, and ineffective against evolving ransomware variants~\cite{DBLP:conf/dimva/KharrazRBBK15}.
Modern ransomware detection solutions have shifted towards dynamic behavioral analysis, which includes real-time monitoring of suspicious file operations patterns~\cite{DBLP:conf/acsac/ContinellaGZPBZ16}, network traffic~\cite{DBLP:journals/csur/OzALU22}, API calls~\cite{DBLP:conf/raid/MehnazMB18}, and registry activities~\cite{hou2024empirical}. By combining machine learning and deep learning techniques, these methods~\cite{DBLP:conf/icacci/SheenY18, DBLP:journals/jnca/AhmedKHAH20, DBLP:journals/compsec/FernandoK22, DBLP:conf/ccwc/MasumFSQ0A22} offer a more robust mechanism against evasion techniques. However, their effectiveness in industrial practices is still tampered by several intrinsic limitations in time constraints~\cite{urooj2022ransomware} and heavy system overhead~\cite{cen2024ransomware,DBLP:conf/raid/MehnazMB18}.
Recently, deception-based detection~\cite{DBLP:journals/tifs/GanfureWCS23,DBLP:conf/raid/MehnazMB18,DBLP:conf/racs/LeeLH17,DBLP:journals/compsec/Gomez-Hernandez18} has been considered a promising way to identify and mitigate ransomware attacks in real-time, offering a proactive defense mechanism to lure, detect, and analyze ransomware activities. These decoys are designed to mimic real system assets~(e.g., project codes or sensitive data), attracting attackers while minimizing interaction with actual resources. 

\noindent \textbf{Research Gap.} Despite these promising results, two substantial gaps remain in translating these advancements into practical solutions in industrial environments: \textit{intolerable system overheads} and \textit{notorious alert fatigue}. 

On one hand, advanced behavior-based detection methods necessitate intensive monitoring and continuous analysis. While theoretically effective for batch processing, these methods require extensive computing resources and result in intolerable delays in industrial-scale streaming data processing. For example, ShieldFS~\cite{DBLP:conf/acsac/ContinellaGZPBZ16} collects runtime data over 60 minutes, resulting in approximately 500MB and nearly 7 million I/O Request Packet~(IRP) logs~(Note that the scale of data encountered in actual industrial settings typically ranges from 10 to 100 times greater than this experimental setup~\cite{splunk2023speed}), and then analyzes in batches for model prediction. However, this batch-processing approach does not align well with the requirements of real-time detection needed in industrial settings. Different from the controlled and post-mortem analysis in academic studies, industrial applications necessitate the streaming of IRP log data directly into detection models to ensure timely threat identification and response. 
According to the latest research from MarauderMap~\cite{hou2024empirical}, behavior pattern monitoring approaches have an average response time ranging from 22.27 to 55.91 seconds.
Even with the most advanced computational resources such as NVIDIA BlueField DPUs~(Data Processing Units) and GPUs~(Graphics Processing Units), the processing time still needs approximately 12 seconds~\cite{nvidia2023ransomware}, which means that up to 20\% of user files could be encrypted by ransomware~\cite{splunk2023speed,nvidia2023ransomware}, highlighting a significant gap in response time that could lead to considerable data loss and system compromise. Moreover, the continuous operation of logging IRP data results in an average user-perceive overhead of 26\%, let alone intensive running inference models. 

On the other hand, the issue of alert fatigue becomes particularly acute in industrial practices. The deployment of decoy files inevitably leads to a high volume of alerts, many of which may be false positives~(FPs). In industrial contexts, the effectiveness of a detection system is significantly compromised by the inability to efficiently manage and prioritize alerts. For example, according to continuous monitoring by RTrap~\cite{DBLP:journals/tifs/GanfureWCS23} over four weeks, the ratio of user files to decoy files to FPs was 510:15:1. This implies that if an industrial environment has a cluster of 1,000 hosts to manage, each with 100,000 user files~(about 54GB according to Splunk Report~\cite{splunk2023speed}), there would be over 196,000 alerts monthly, an overwhelming burden for security analysts~\cite{risk2021avoiding}. Faced with this challenge, EDR solutions in practice often resort to directly terminating suspicious processes without manual checks, potentially disrupting critical business or system operations. To avoid such disruptions, industrial practitioners have to whitelist critical business or system processes. This coarse-grained approach, however, renders the EDR system ineffective against ransomware variants that employ whitelist-process injection techniques~(about 39.1\%~\cite{hou2024empirical}). Striking the right balance between false positives and false negatives remains an ongoing research gap in the industrial implementation of ransomware detection systems.

To bridge the gap between academic research and industrial implementation of ransomware detection, we propose \toolname\footnote{The name is derived from the metaphorical phrase ``\underline{\textbf{Can}}ary \underline{\textbf{Cal}}ls'', signifying its role as an early warning system against ransomware threats, akin to the canary's distress calls in coal mines.}, a novel \textbf{Monitoring-Detection-Response~(MDR)} framework that combines \textit{lightweight monitoring}, \textit{selective detection}, and \textit{real-time response} to enable effective and efficient protection against evolving ransomware threats. Specifically, \toolname employs a two-pronged canary monitoring strategy that synergistically integrates decoy files and ransom note monitoring. Instead of indiscriminately logging IRP for all processes, \toolname focuses on processes that trigger the monitoring points. By performing an in-depth analysis of the IRP logs generated by the suspicious processes reported at the monitoring layers, \toolname can detect unknown threats almost in real time and combat alert fatigue caused by decoy files. We next describe the technical challenges and key insights in bridging these research gaps.

\noindent \textbf{Technical Challenges.} Implementing such a lightweight and real-time detection system presents three technical challenges:

\begin{itemize}[noitemsep, topsep=1pt, partopsep=1pt, listparindent=\parindent, leftmargin=*]
    \item[\ding{172}] \textit{How to achieve real-time ransomware detection with lightweight behavioral monitoring and model inference while minimizing system overhead?} The volume and velocity of data generated in industrial-scale environments pose significant challenges for real-time detection and response. Traditional batch processing approaches~\cite{DBLP:conf/acsac/ContinellaGZPBZ16,DBLP:conf/uss/KharrazAMRK16,DBLP:conf/raid/KharrazK17,DBLP:conf/icdcs/ScaifeCTB16}, which collect runtime data over extended periods and analyze them offline, are not suitable for the timely detection and mitigation of ransomware attacks. The delays introduced by batch processing can allow ransomware to encrypt a significant portion of user files before response actions are triggered, leading to substantial data loss.

    \item[\ding{173}] \textit{How to ensure monitoring points capture all potential ransomware processes with the highest possible recall rate?} Ransomware employs sophisticated evasion techniques to bypass detection, making it challenging to capture all potential ransomware processes. For example, ransomware may selectively encrypt files based on certain criteria~(e.g., file type, size, or location) to avoid triggering decoy file~\cite{DBLP:journals/tifs/GanfureWCS23,DBLP:conf/cns/DenhamT23,urooj2022ransomware}, resulting in false negatives. Improving the recall rate of monitoring points is a daunting challenge that requires a multi-faceted approach.

    \item[\ding{174}] \textit{How to effectively identify and isolate the ransomware process, akin to finding a needle in a haystack of massive alerts triggered by monitoring points?} The vast number of alerts generated by monitoring points can overwhelm security analysts, leading to alert fatigue and missed detections. Reducing false positives and accurately identifying the ransomware process among the sea of alerts is crucial for effective response and mitigation. 
\end{itemize}

\noindent \textbf{Key Insights.} To address these challenges, \toolname incorporates several key insights and techniques:

\begin{itemize}[noitemsep, topsep=1pt, partopsep=1pt, listparindent=\parindent, leftmargin=*]
    \item \textbf{Lightweight Process Filtering.} To achieve real-time detection with minimal performance overhead, \toolname employs a lightweight process filtering mechanism. Instead of indiscriminately monitoring all processes, \toolname focuses on suspicious processes that exhibit behaviors indicative of ransomware activity. By selectively capturing critical system events and behaviors associated with these suspicious processes, \toolname significantly minimizes storage requirements and computational overhead, allowing for real-time detection and response.

    \item \textbf{Integrated Canary Monitoring.} \toolname addresses the challenge of capturing all potential ransomware processes by employing an integrated canary monitoring approach. Unlike traditional decoy-based strategies that rely solely on decoy files, \toolname combines multiple monitoring methods, including decoy file access monitoring and ransomware note monitoring. The integration of diverse monitoring methods provides a more comprehensive and resilient detection mechanism, improving the recall rate of capturing ransomware processes.

    \item \textbf{Fine-grained Detection.} To effectively identify and isolate ransomware processes among the vast number of alerts triggered by monitoring points, \toolname employs fine-grained detection techniques. \toolname leverages the gradient-boosting decision tree to analyze the behavioral patterns and characteristics of suspicious processes, enabling accurate differentiation between benign and malicious activities and minimizing alert fatigue.
\end{itemize}

\noindent \textbf{Contributions.} Based on these insights, \toolname develops three key components to address these challenges:

\noindent\emph{(1) Lightweight Decoy File Monitor.} 
We introduce a novel approach that transitions from passive damage acceptance to active ransomware enticement. By strategically placing lightweight decoy files as traps, \toolname lures ransomware into exposing itself during the early stages of infiltration. This enables real-time interception and minimizes the impact on system performance, effectively addressing the challenge of early detection.

\noindent\emph{(2) Semantic-based Ransom Note Monitor.} 
We leverage the distinct textual patterns present in ransom notes to identify ransomware infections. By performing semantic analysis on newly created or modified files, \toolname swiftly detects ransom notes and alerts the victim. This approach overcomes the potentially bypassable limitations of decoy-based monitoring and allows behavioral detection of specific processes to be triggered.

\noindent\emph{(3) Multi-granularity Behavior Detector.} 
We develop an advanced decision engine that combines automatically constructed embedding with manually collected expert features to capture ransomware's complex and evolving behavior. By representing processes' behavior as bipartite graphs and encoding patterns into compact embeddings, the detector learns intricate behavioral characteristics. The integration of expert knowledge further enhances the decision-making process. This multi-granularity approach effectively addresses the challenge of detecting sophisticated and evolving ransomware variants.

\noindent \textbf{Evaluation.} 
We construct an automated simulation system and collect a large-scale sample library of behavior logs, encompassing 3 million event sequences from 1,335 known ransomware, 1,768 benign software, and 426 unknown samples capable of evading anti-virus engine detection. The experimental results indicate that \toolname achieves a true positive rate of 99.65\% on a large dataset comprising 555,678 unknown ransomware behaviors, with near-zero false positives. Not only does \toolname demonstrate exceptional detection capabilities, but it also exhibits efficient real-time response and a lightweight design. Our ablation study reveals that, compared to traditional approaches relying solely on decoy monitoring or file behavior monitoring, \toolname's multi-layered detection mechanism significantly reduces resource consumption and false positive alerts. For instance, the suspicious process filtering based on monitoring points reduced the average CPU utilization by 91.04\% ($6.7\% \to 0.6\%$) and the peak CPU utilization by 76.69\% ($26.6\% \to 6.2\%$), while achieving rapid model inference within 30ms and real-time response within 3 seconds. Furthermore, continuous tests over 9 months demonstrated that the file behavior detector effectively mitigates the overwhelming number of alerts triggered by monitoring points~($3,192 \to 750$), avoiding 76.50\% of the unexpected process interruption and required inspection efforts from security analysts.

By the time of this writing, \toolname has been integrated into a commercial product and successfully deployed on 3.32 million endpoints for over a year. From March 2023 to April 2024, \toolname has successfully detected and thwarted 61 ransomware attack incidents spanning various industries, including manufacturing, technology, services, and transportation. \toolname can effectively detect both well-known and distinctive variants of ransomware, including Mallox, Lockbit 3.0, and Tellyouthepass. Notably, according to a manual forensic of 27 ransomware detected from March to June 2023, \toolname has defended against 13 n-day attacks that exploited vulnerabilities documented in the CVE database and prevented 5 high-risk zero-day attacks, further demonstrating its ability to combat unknown and sophisticated ransomware threats. Overall, \toolname is highly effective in defending against real-world ransomware attacks on user endpoints and can provide reliable protection for various industries.
\section{Design Details of \toolname}

\begin{figure}[t] 
    \centering 
    \includegraphics[width=0.47\textwidth]{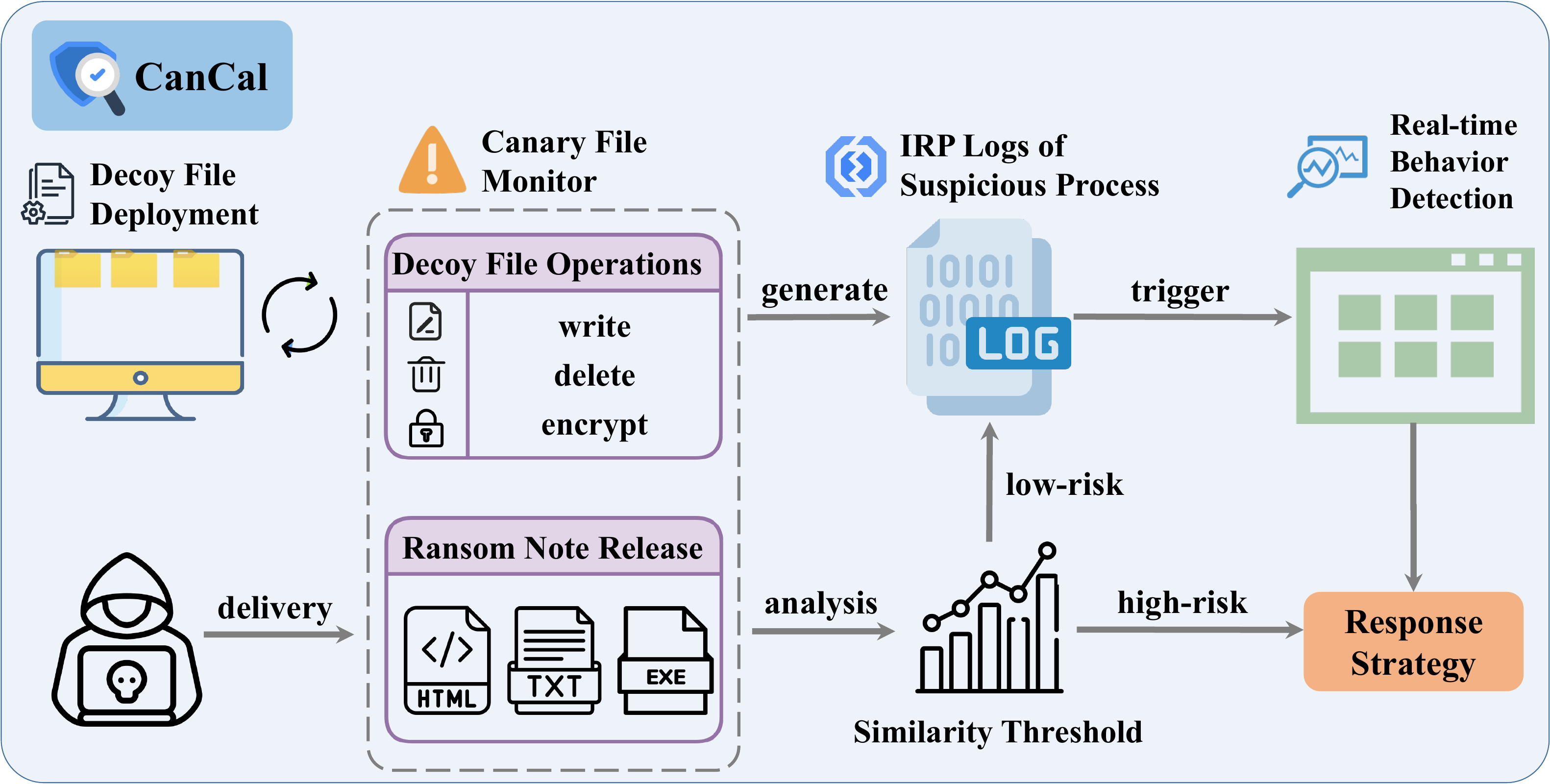} 
    \caption{Overall architecture of \toolname.}
    \label{fig1}
\end{figure}

This section introduces the framework of ``Monitoring-Detection-Response'' and the layered architecture of \toolname. The full workflow is shown in \autoref{fig1}.

\noindent \textbf{Monitoring:} This process involves two components:
(1)~\textit{Decoy File Monitor} deploys decoy files and closely monitors suspicious activities like deletion, writing, or encryption. If detected, further analysis and detection are performed.
(2)~\textit{Ransom Note Monitor} detects the execution of ransomware. When the ransomware executes, it releases a ransom note. \toolname analyzes the similarity of ransom notes and assigns a threat level accordingly.

\noindent \textbf{Detection:}
The detection component aims to identify ransomware processes from the vast number of low-risk level threat alerts reported by monitoring points. This engine collects IRP logs of suspicious processes and combines expert-based features with automatic embedding to capture the ransomware behavior pattern.

\noindent \textbf{Response:}
For low-risk threat levels, \toolname relies on real-time behavior detection to track the suspicious process.
While for high-risk threat levels, \toolname takes appropriate response measures to mitigate the impact.
This may involve terminating the process and isolating the endpoint.

\subsection{Lightweight Decoy File Monitor}
\toolname proposes a real-time dynamic detection method based on decoy files, which serves as the detection trigger layer to enable the transition from passive acceptance of ransomware damage to active enticing. 
By deploying carefully crafted decoy traps, \toolname continuously monitors system activities and implements counter-interception during the ransomware invasion phase.

\subsubsection{Decoy File Design}
In the context of ransomware detection, the primary objective of decoy file design is to enhance their credibility and access priority, thereby ensuring their interaction with ransomware. A suboptimal design may lead to two undesirable outcomes: either the ransomware identifies and circumvents the decoy, compromising detection accuracy, or it prioritizes the encryption of legitimate files before accessing the decoy, thus impairing the timeliness of detection. To address these challenges, we have developed a holistic approach to decoy file design, encompassing file names, types, and content.

$\bullet$ \textit{Decoy File Name and Type:} To effectively deceive ransomware, the naming convention for decoy files must be meticulously crafted. Our approach aims to mimic the nomenclature of legitimate files as closely as possible, minimizing the risk of detection by ransomware. We prioritize information-rich file types for our decoys, such as documents and images, which are typically high-value targets for ransomware attacks.

$\bullet$ \textit{Decoy File Content:} To bolster the credibility of our decoys, we ensure that their content is contextually appropriate for their purported file type. For instance, document-type decoy files are populated with plausible, randomly generated text that aligns with the expected content of such files.

\subsubsection{Decoy File Deployment}
\toolname is engineered to provide real-time protection for end users by leveraging a simulated environment that closely mimics genuine hardware and software systems. This approach is designed to provoke potential malicious activities and enhance interaction with ransomware threats, thereby significantly reducing the likelihood of evasion. The system employs two distinct deployment strategies: automatic deployment and user-defined placement.

$\bullet$ \textit{Automatic Deployment:} This method involves a thorough analysis of typical ransomware file traversal patterns, with particular attention to the initial directories accessed during these traversals. Common targets, such as \verb*|C:\Users\username\AppData|, are identified and utilized as predefined locations for decoy file placement. These strategically positioned decoys are then subject to real-time monitoring. This approach enables the endpoint to swiftly detect any modification attempts and implement immediate countermeasures to prevent ransomware propagation. By ensuring early access to decoy files, this method substantially enhances overall system security.

$\bullet$ \textit{User-Defined Placement:} This technique empowers users to customize the location of decoy files, thereby offering enhanced protection for critical data. Users simply specify the desired number of decoys and their intended locations within the file system. This method strikes an optimal balance between the security benefits of manual decoy placement and the efficiency of automated systems, significantly reducing the time required for internal system maintenance.

\subsubsection{Decoy File Monitoring}

Upon any modification to a decoy file, the system's callback function initiates a filtering interception logic. This triggers an alert within the system's trigger layer, signaling a potential security breach. To mitigate false positives and ensure accuracy, the system doesn't immediately classify this as a ransomware attack. Instead, it prompts an in-depth ransomware screening process, activating targeted process monitoring and comprehensive behavior collection for the suspect processes. The collected data is then analyzed by \toolname's sophisticated dynamic behavioral engine for definitive threat discrimination.

\begin{figure}[t] 
    \centering 
    \includegraphics[width=0.45\textwidth]{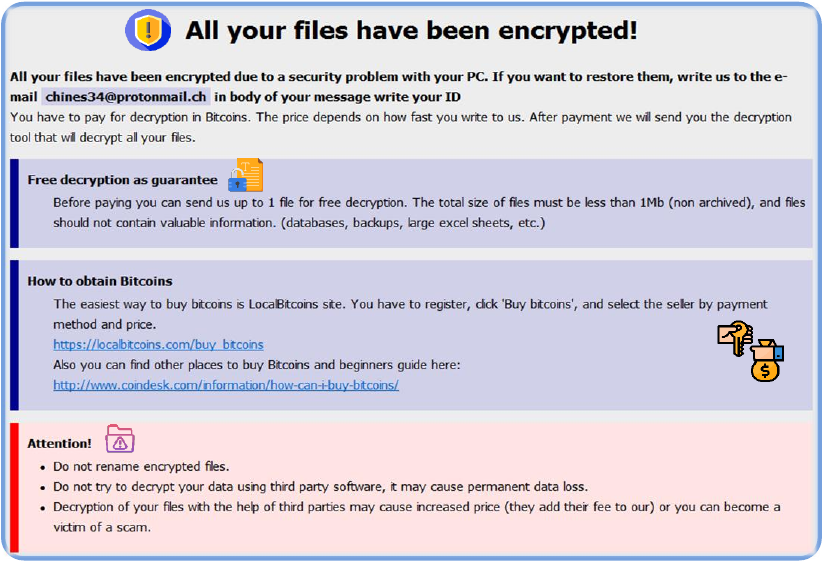} 
    \caption{An example of the ransom note.}
        \label{fig2}
\end{figure}

\subsection{Semantic-based Ransom Note Monitor}
Once files are encrypted or devices are disabled, ransomware attackers typically alert victims to the infection.
The usual practice is to use a pop-up notification or release a file (txt/html/exe, etc.) deposited in a file directory or on the computer desktop. 
\autoref{fig2} illustrates a typical ransom note issued by ransomware, wherein victims are instructed to remit payment in cryptocurrency to recover their files.
These ransom notes generally contain comprehensive information about the ransomware attack, including encryption status, ransom amount, and payment methods, often presented in a consistent textual pattern. Leveraging this consistency, semantic analysis of ransom notes can effectively detect the behavioral characteristics of ransomware in real-time, thereby enhancing preventive measures against such attacks.

\subsubsection{Ransom Note Content}
Ransom notes typically delineate the crypto-extortion scenario and outline the payment process, usually demanding cryptocurrency or similarly untraceable methods in exchange for data recovery and system restoration. The fundamental components of these notes generally include:

$\bullet$ \textit{Headline:} Usually an eye-catching headline like ``Your files are encrypted'' or ``Your computer is locked'' is employed to attract the victim's attention.

$\bullet$ \textit{Threat:} The ransom note usually details the compromised state of the victim's computer or files, emphasizing their encrypted or locked status. It often warns that failure to pay the ransom will result in permanent inaccessibility or corruption of files. Additionally, threats of exposing the victim's personal information and data are common tactics.

$\bullet$ \textit{Payment Information:} Attackers typically include a payment link or cryptocurrency address, along with detailed instructions for submitting the ransom.
 
$\bullet$ \textit{Deceptive Offers:} To enhance their perceived credibility and manipulate victims, attackers may present ostensibly benevolent options such as ``free decryption'' or ``file recovery'' services.

$\bullet$ \textit{Contact Information:} To further their deception, attackers often provide contact details, such as an email address, under the guise of offering support or facilitating ransom payment.

\subsubsection{Ransom Note Analyzer}

According to the basic composition of ransom notes, it can be found that most of the expressions revolve around the specific behavior of ransomware attacks, such as ``encryption'', ``decryption'', ``file loss'', etc. Using natural language processing technology, it is possible to automate the recognition and matching of repetitive language structures and vocabulary. This technology plays an important role in the real-time monitoring of ransomware, enabling the immediate identification of potential threats at the moment of ransom note release.
Specifically, this module can be divided into three steps: parsing of ransom notes, gene pool creation, and similarity discrimination.

\noindent \textbf{Step 1: Ransomware Parser.} To capture the context of encryption and ransom, this module is designed with a parsing step. 
This parsing can divide the long ransom text into small unit fragments for analysis.
Specifically, the word-level $n-gram$-based method is used to construct a local continuous sequence. 
Given a ransom note of length $k$, $W =\left(w_1, w_2,w_3, \cdots,w_k \right)$ , where $w_i$ denotes the $i$-th word in the text segment $W$.
A window of size $n$ is slid over this text segment, and the $n$ words in the current window are selected as segments at each move.
In this way, the set of all $n-gram$ sequences are extracted, consisting of $W_{ngram} = \left\{(w_1, w_2, \cdots, w_{n}), \cdots, (w_{k - n + 1}, w_{k - n + 2},\cdots, w_{k}) \right\}$, where the set size is $ k - n + 1$, each sequence $(w_{i}, w_{i+1}, \cdots, w_{i + n - 1})$ consists of the adjacent $n$ words in the text segment $W$.

\begin{figure}[!t] 
    \centering
    \includegraphics[width=0.45\textwidth]{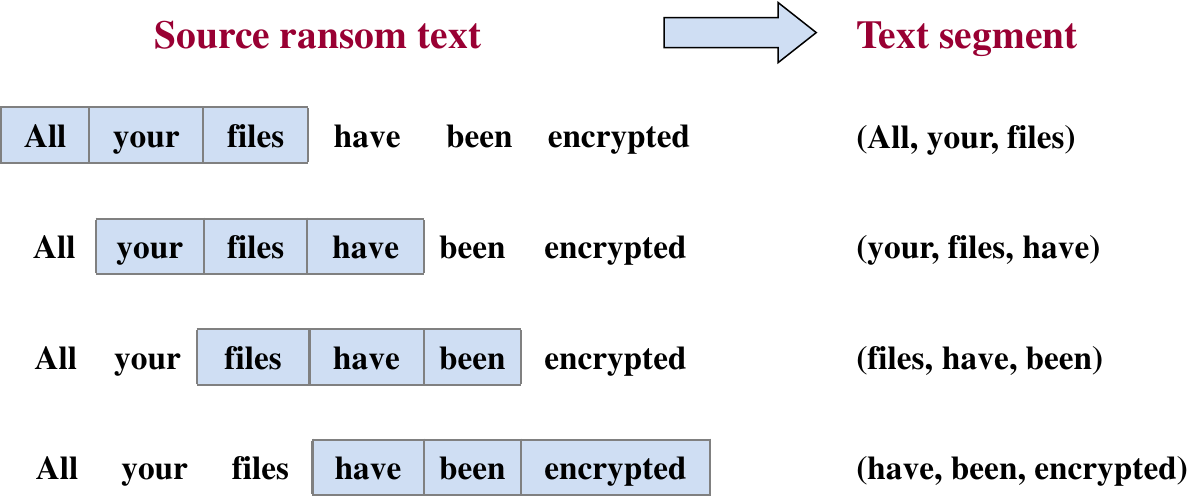} 
    \caption{Sliding window of trigram method. It extracts three consecutive text words in the ransom note.}
    \label{fig3}
\end{figure}

In the \textit{n-gram} method, the larger the selected value of $n$, the richer the contextual information that the model can capture, but it also leads to the problems of subsequent matching errors. 
After analyzing a set of real ransom note samples, we have determined that the optimal $n$ value for our system is 3
~(See \crappendix{A}).
It strikes a balance between capturing contextual information and avoiding subsequent matching errors. 
While larger $n$ values can capture richer contextual information, they also may be more sensitive to minor variations in phrasing, which could result in missed matches for semantically similar but syntactically different expressions.
Conversely, smaller $n$ values such as unigrams or bigrams may not provide sufficient context for accurate recognition of ransomware-specific language patterns.
Specifically, \autoref{fig3} shows an example, which involves extracting sequences of three consecutive words.
When considering three adjacent words, the consecutive sequences ``all your files'' and ``have been encrypted'' are extracted.
They appear frequently as important linguistic features of ransom notes and can express more complete ransom semantics. 
Therefore, we choose $n=3$ as the optimal value to improve the ransom note classification accuracy.

\noindent \textbf{Step 2: Create a Ransomware Family Gene Pool.}
To analyze the textual features of ransom notes, we conduct statistical analysis on each parsed text sequence $(w_{i}, w_{i+1}, w_{i + 2})$. Specifically, we quantify the frequency of occurrence $c_i$ for each sequence within the entire set, as frequently appearing sequences likely represent key information in the ransom note. We then normalize these frequency counts to scores $f_i$, which are arranged in descending order. This process allows us to compare the relative importance of different word sequences and identify the most prominent linguistic patterns. This normalization process is mathematically expressed as:
\begin{equation}
    f_i = \frac{c_i}{\sum_{m = 1}^{k - n + 1}{c_m}}
\end{equation}

Based on these key fragments, we further construct a gene pool of ransomware families for storing and managing the information in ransom notes. This gene library contains all common key fragments and their normalized values in different ransom notes. This curated collection of linguistic features enhances our ability to efficiently identify and detect potential ransom notes.

\begin{figure*}[!t] 
    \centering
    \includegraphics[width=\textwidth]{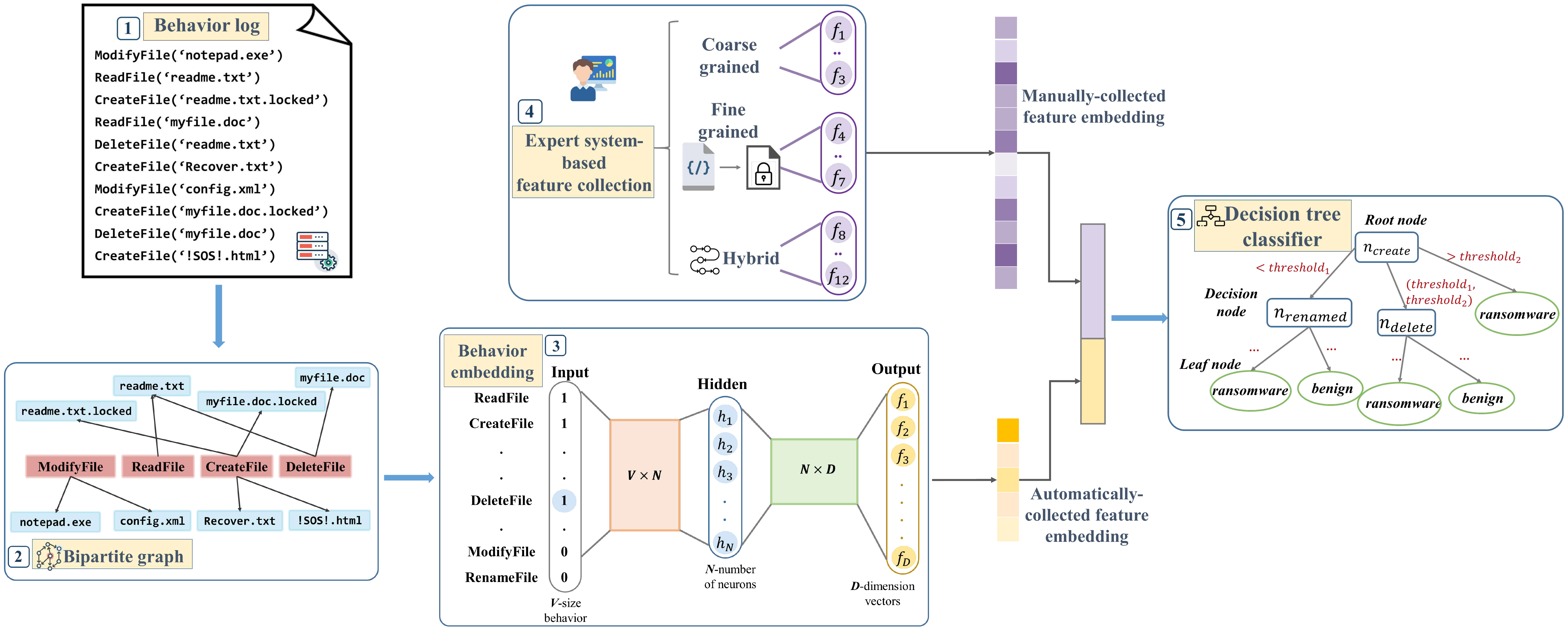} 
    \caption{An overview of the behavioral engine with multi-granularity features. The engine includes five main modules:
(1) Behavior log collection module: records the process activities;
(2) Behavior graph construction module: converts the behavioral operations into a two-part "instruction-parameter" behavior graph;
(3) Behavior pattern encoding module: encodes each behavioral operation into an embedded representation;
(4) Expert knowledge feature module: extracts features based on expert experience;
(5) Classifier module: performs ransomware classification using the decision tree algorithm.
    }
    \label{fig4}
\end{figure*}

\noindent \textbf{Step 3: Calculate Ransom Note Similarity.} Once the ransomware gene fragments are constructed, we can determine whether a given file $d$ exhibits characteristics of ransom note by comparing its similarity to these established gene fragments.
Specifically, the sequence fragments are extracted from the file using the same parsing method. Then, we calculate the similarity score by summing the normalized values $f_i$ of the $t$ gene fragments that successfully match:
\begin{equation}
    sim(d) = \sum_{i = 1}^{t}f_i
\end{equation}

To determine whether the current sample is a ransom note, we compare its similarity score to a threshold $\tau_{sim}$. This threshold is established by analyzing the similarity of known samples to the gene pool of ransomware families. A detailed analysis of the similarity threshold and its determination can be found in 
\crappendix{A}.

\subsection{Multi-granularity Behavior Detector}
The behavioral engine is an advanced AI-powered tool designed to enhance the analysis and classification of ransomware.
By leveraging both automatically collected logs and manually constructed features, this engine significantly improves the performance of the ransomware classifier. The overall architecture is shown in \autoref{fig4}, and the specific implementation consists of the following steps:

\noindent \textbf{Step 1: Collect Behavior Logs.} \toolname automatically collects process behavior logs, which include the sequence of system events during execution and their corresponding parameters. These logs provide detailed information about the behavior of each process.

\noindent \textbf{Step 2: Construct Behavior Graphs.} \toolname converts the behavioral operations from the logs into bipartite behavior graphs. These graphs contain two types of vertices: event operations and parameters. By drawing edges between each behavioral operation and its parameters, the graph structure enables better extraction of specific behavioral patterns~(e.g., network communication, file encryption modifications, etc.).

\noindent \textbf{Step 3: Encode Behavior Patterns.} Each behavior pattern is compressed into a sparse one-hot vector, which indicates whether a specific operation or parameter is present in each process. A neural network then converts this vector into a compact pattern embedding, facilitating the learning of more complex feature representations and enhancing the performance of the ransomware classifier.

\noindent \textbf{Step 4: Combine Expert Knowledge.} \toolname combines the automatically collected behavior embeddings with multi-granularity behavior features derived from the expert system. These combined features serve as the final input to the decision tree classifier.

\noindent \textbf{Step 5: Train a Decision Tree Model.}
Using the collected data and integrated features, \toolname trains a decision tree classifier to determine whether a process is ransomware or not. The decision tree classifier autonomously selects the optimal path for classification based on the input feature vector.

Note that the expert knowledge features were derived from the experience of an industrial ransomware emergency response team, comprising 6 members with an average of 6-8 years of experience. The selection process followed a rigorous approach, involving independent proposals, structured discussions, and consensus voting. The following section will delve into the detailed design and rationale behind these expert-derived features.

\begin{figure*}[!t] 
	\centering 
	\includegraphics[width=\textwidth]{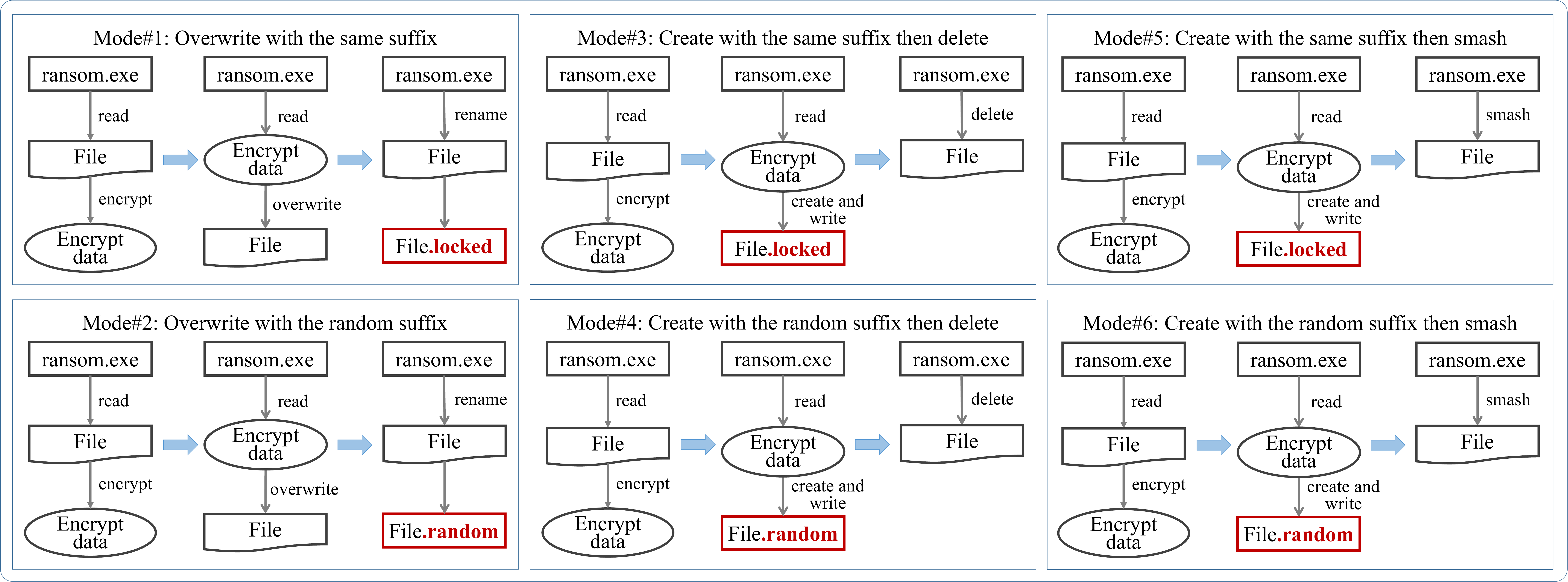} 
	\caption{There are six distinct encryption modes, each characterized by different I/O access modes and encryption suffix names:
(1) The attacker creates an encrypted file with the same suffix name and overwrites the original file.
(2) This mode is distinguished from Mode\#1 by creating encrypted files with random suffixes.
(3) The attacker creates an encrypted file with the same suffix name and deletes the original file.
(4) This mode is distinguished from Mode\#3 by creating encrypted files with random suffix names.
(5) The attacker creates an encrypted file with the same suffix name and smashes the original file.
(6) This mode is distinguished from Mode\#5 by creating encrypted files with random suffix names.
}
\label{fig5}
\end{figure*}

\subsubsection{Ransomware Encryption Mode}
To reach the goal of hiding itself, encrypting victim host files, and demanding ransom, ransomware usually performs frequent operations on files and presents significant anomalous features. To address this issue, we focus on file encryption behavior, and according to different I/O access policies, existing file encryption methods can be divided into two main categories \cite{DBLP:conf/uss/KharrazAMRK16,zhou2023limits}: (1) Overwrite: The ransomware reads the file content and encrypts it, then writes it back to the original file directly; (2) Create and Delete: The ransomware creates a new ciphertext and then deletes the original text; (3) Create and Smash: The ransomware creates a new ciphertext and then smashes the original text with random data.

\subsubsection{Expert Knowledge Feature Engineering}
\label{feature}
The essence of feature engineering is to extract the maximum number of features for use in the model. Taking the encryption of important files as an example, different encryption strategies use different file operations, which can be characterized by analyzing the involved file operations for behavioral characterization.

The process is the basic unit of ransomware scheduling operations. When the suspicious process $p$ triggers the monitoring point $m$, the file operations are recorded during the duration period $\Delta t$. Specifically, the operation of each process consists of elements such as timestamp, process number, and file name, in the form of a five-tuple, \verb*|<Time,Pid,Pid_name,Operation,File_name,| \verb*|File_type>|. Based on this information, the number of file changes in the monitored time period is counted, and the characteristics of different levels are fused to build a feature set of file operations. This feature set can be specifically divided into three categories:

\noindent\textbf{Coarse-grained Features Based on File Operations.} 
This type of feature is mainly based on the operation behavior of file creation, file deletion, file renaming, etc. We count the number of file changes in process-based units.
Specifically, by monitoring the behavior of processes in the system, the number of changes over a period of time is recorded and counted. The following are the definitions of this type of feature:
\begin{mydef}
Number of files created $n_{create}$. Ransomware creates a large number of files to store ransom notes or encrypted.
\end{mydef}

\begin{mydef}
Number of deleted files $n_{delete}$. Ransomware deletes the original files during encryption to prevent users from recovering them.
\end{mydef}

\begin{mydef}
Number of renamed files $n_{renamed}$. The ransomware renames or modifies the extension of files during encryption.
\end{mydef}

\noindent\textbf{Fine-grained Features Based on Different Encryption Modes.}
During the encryption process, attackers often add or modify file suffixes, resulting in changes to the file type that make it difficult to access.
In ransomware-initiated encryption, suffix modifications can occur in two scenarios: first, by adding the same file suffix to all target files, such as the name of the ransomware family; and second, by randomly replacing the file suffix with a different string. To account for the different characteristics of I/O modes and suffix names, this study divides encryption operations into six modes, which are defined as follows.
\autoref{fig5} illustrates the six modes of encryption operations.

$\bullet$ Mode\#1: In this mode, the original file is overwritten with the encrypted file, and the same suffix name is added uniformly to all affected files.

$\bullet$ Mode\#2: This mode involves the encrypted file overwriting the original file, with different suffix names added randomly during the encryption process.

$\bullet$ Mode\#3: This mode involves deleting the original file and creating a new encrypted file, with the same suffix name added uniformly to all affected files.

$\bullet$ Mode\#4: This mode also involves deleting the original file and creating a new encrypted file, but with different suffix names added randomly during the encryption process.

$\bullet$ Mode\#5: This mode involves smashing the original file and creating a new encrypted file, with the same suffix name added uniformly to all affected files.

$\bullet$ Mode\#6: This mode also involves smashing the original file and creating a new encrypted file, but with different suffix names added randomly during the encryption process.

When the process $p_i$ triggers any monitoring point, the number of file types $ntype,ntype'$ before and after the monitoring time period $\Delta t$ are recorded respectively. To further analyze and identify different encryption behavior features, we count the increase or decrease in the number of file types. Although six encryption modes are described, from the perspective of $ntype_{change}$, these modes actually result in four distinct features, which can be calculated as:
\begin{equation}
	ntype_{change} = ntype' - ntype
\end{equation}

\noindent\textbf{Fusion-based Features.} 
We employ a multi-feature fusion approach to leverage the complementary advantages of diverse features, which enhances the accuracy and reliability of our feature set, thereby improving ransomware classification capabilities.
\begin{mydef}
Ratio of file types ($rtype$): 
This feature represents the relative change in file types before and after the monitoring period. Let $ntype$ and $ntype'$ denote the number of file types before and after the monitoring period, respectively.
\end{mydef}

\begin{equation}
    rtype = \frac{ntype'}{ntype}
\end{equation}

\begin{mydef}
Ratio of deleted to created file types ($rtype_{change}$): This feature aids in identifying ransomware with different encryption modes by analyzing file type changes.
\end{mydef}
During the monitoring period $\Delta t$, we record $ntype_{del}$ (the number of deleted file types) and $ntype_{create}$ (the number of created file types). The ratio is then computed as:

\begin{equation}
    rtype_{change} = \frac{ntype_{del}}{ntype_{create}}
\end{equation}

\begin{mydef}
Maximum number of created files with the same name ($max(n_{file})$): This feature helps identify ransomware behavior, as these malicious programs often distribute multiple ransom notes with identical filenames.
\end{mydef}

During $\Delta t$, we count the occurrences $n_{file}$ of each unique filename and determine the maximum value $max(n_{file})$.

\begin{mydef}
Number of folders containing the most files with the same name ($n_{folder}$): This feature distinguishes ransomware behavior from normal file operations.
\end{mydef}

For files whose creation count reaches $max(n_{file})$, we analyze their distribution across different folders during $\Delta t$. While benign software typically creates files with the same default filename in a single directory, ransomware often distributes identical ransom notes across multiple directories.

\begin{mydef}
Ratio of file creation with the same name ($r_{file}$): This feature combines the above two metrics to provide a comprehensive view of file creation patterns.
\end{mydef}
    \begin{equation}
        r_{file} = \frac{max(n_{file})}{n_{folder}}
    \end{equation}

Further insights into the design of these expert knowledge features are provided in 
\crappendix{B}.

\begin{algorithm}[t]
\SetAlgoLined
\DontPrintSemicolon
\SetKwFunction{TreeGenerate}{TreeGenerate}
\SetKwProg{Fn}{Function}{:}{end}
\fontsize{8.5}{10}\selectfont
\caption{\textbf{Decision Tree Generation}}
\label{alg:decision_tree}

\KwIn{Training dataset $D={(x_1, y_1),(x_2, y_2), \ldots,(x_m, y_m)}$,\
Feature set $A = {a_1, a_2, \ldots, a_n}$}
\KwOut{Decision tree $f(x)$}

\BlankLine
\Fn{\TreeGenerate{$D$, $A$}}{
    $node \leftarrow \text{new Node}()$\;
    \If{$\forall i \in \{1,2,\ldots,m\}: y_i = c$}{
        $node.class \leftarrow c$\;
        \Return $node$
    }
    \If{$A = \emptyset \lor \forall i \in \{1,2,\ldots,m\}: x_i = T$}{
        $node.type \leftarrow \text{leaf}$\;
        $node.class \leftarrow \arg\max_{c \in C} \sum_{i=1}^n [y_i=c]$\;
        \Return $node$
    }
    $a_* \leftarrow \arg\max_{a \in A} \text{InformationGain}(D, a)$\;
    \ForEach{$a_*^v \in \text{Domain}(a_*)$}{
        $child \leftarrow \text{new Node}()$\;
        $D_v \leftarrow \{ (x, y) \in D : x_{a_*} = a_*^v \}$\;
        \eIf{$D_v = \emptyset$}{
            $child.type \leftarrow \text{leaf}$\;
            $child.class \leftarrow \arg\max_{c \in C} \sum_{i=1}^{n} [y_i = c]$\;
        }{
            $child \leftarrow$ \TreeGenerate{$D_v, A \setminus \{a_*\}$}\;
        }
        $node.\text{addChild}(a_*^v, child)$\;
    }
    \Return $node$
}
\end{algorithm}

\subsubsection{Attack Detection Classifier}
In large-scale industrial environments, we exclude computationally demanding deep learning approaches due to performance constraints and limited GPU resources. Adopting a traditional machine learning model was a pragmatic decision.
In practice, \toolname leverages the Gradient Boosting Decision Tree~(GBDT) algorithm~\cite{Friedman2001GreedyFA} for the attack detection process, specifically implemented using XGBoost~\cite{chen2016xgboost}~(the comparative experiment details are provided in 
\crappendix{C}).
By combining the advantages of both decision tree~\cite{DBLP:books/wa/BreimanFOS84} and gradient boosting models~\cite{DBLP:conf/ausdm/SteinbergGC02}, a classifier for ransomware is built.
Following the concept of ensemble learning, a group of weak learners are integrated into a strong learner.
When classifying benign software and ransomware, the decision tree comprises nodes and directed edges, where internal nodes represent features and leaf nodes represent software categories.
The main principle of the GBDT approach is to use gradient descent to approximate each decision tree. Specifically, each iteration minimizes training errors along the negative gradient direction.

Consider a training set $D=\left\{\left(x_1, y_1\right),\left(x_2, y_2\right),\cdots,\left(x_m, y_m\right)\right\}$ comprising $m$ samples, where $x_i=\left(x_i^1, x_i^2. \cdots, x_i^n\right)^{\mathrm{T}}$ represents the $n$-dimensional feature vector of the $i$-th sample, and $y_i \in\{0,1\}$ denotes its corresponding category label.
The decision tree generation process is outlined in \autoref{alg:decision_tree}.
This algorithm partitions the input space into disjoint regions. Specifically, it selects the $j$-th dimensional feature $a_*$ from the $n$-dimensional feature space and determines a criterion value $a_*^v$ for division. Samples are then allocated to either the left subtree $R_{1}(a_*, a_*^v)$ or the right subtree $R_{2}(a_*, a_*^v)$, based on whether their $j$-th dimensional feature value is less than or equal to $a_*^v$.
The training objective is to identify the optimal division feature $a_*$ and its corresponding threshold $a_*^v$ that minimize the prediction error of both subtrees:
\begin{equation}
\label{equ1}
\min _{a_*, a_*^v}\left[\min \sum_{x_i \in R_{1}}\left(y_i-\overline{y}_1\right)^2+\min \sum_{x_i \in R_{2}}\left(y_i-\overline{y}_2\right)^2\right]
\end{equation}
where the divided left subtree is $R_1(a_*, a_*^v) = \left\{x \mid x^{(a_*)} \leq a_*^v\right\}$ and the right subtree is $R_2(a_*, a_*^v) = \left\{x \mid x^{(a_*)}>a_*^v\right\}$. 
The average of the label values of the left subtree $\overline{y}_1=\operatorname{ave}\left\{y_i \mid x_i \in R_{1}(a_*, a_*^v)\right\}$, and the corresponding right subtree $\overline{y}_2=\operatorname{ave}\left\{y_i \mid x_i \in R_2(a_*, a_*^v)\right\}$.
XGBoost extends this objective by incorporating a regularization term $\Omega(f)$ to control model complexity, where $\Omega(f) = \gamma T + \frac{1}{2}\lambda \|w\|^2$, with $T$ representing the number of leaves in the tree and $w$ denoting the leaf weights.

The computation process for selecting optimal division criteria can be further optimized using \autoref{equ2}, reducing the time complexity to $\mathcal{O}(nm^2)$. Notably, the feature dimension $n$ is 12, which is significantly smaller than the number of samples $m$.

\begin{equation}
\label{equ2}
\begin{aligned}
    &\min _{a_*, a_*^v}\left[\min \sum_{x_i \in R_{1}}\left(y_i-\overline{y}_1\right)^2+\min \sum_{x_i \in R_{2}}\left(y_i-\overline{y}_2\right)^2\right]\\
    &= \sum_{x_i \in R}y_i^2 - \min _{a_*, a_*^v}\left(\sum_{x_i \in R_{1}}\frac{1}{m_1}y_i^2+ \sum_{x_i \in R_{2}}\frac{1}{m_2}y_i^2\right)
\end{aligned}
\end{equation}

This optimization approach employs functional gradient descent, applying the steepest descent step to the minimization problem. The objective is to locate a local minimum of the loss function by iterating on the previous function value, denoted as $f_{t-1}(x)$. Consequently, the local maximum-descent direction of the loss function is represented by its negative gradient.
\section{Evaluation}
To comprehensively evaluate \toolname on ransomware detection, we conducted a series of experiments to answer the following research questions:

$\bullet$ \textbf{RQ1~(Effectiveness):} How do the quantitative results of the training process of the \toolname?

$\bullet$ \textbf{RQ2~(Comparison):} How does the \toolname perform in detecting threats on public datasets, and how does its performance compare to other state-of-the-art solutions?

$\bullet$ \textbf{RQ3~(Ablation Study):} How does each component affect the performance of \toolname separately?

$\bullet$ \textbf{RQ4~(Industrial Deployment):} How well does \toolname perform when it is deployed in real-world industrial environments to defend against ransomware attacks?


\subsection{Experiment Setup}

To conduct the experiments of RQ1\string~3, we simulate the operating system environment using virtual sandbox technology. 
The virtual machine is based on the Windows operating system and is equipped with the necessary software and configuration environment. 
To ensure the security and stability of the virtual machine, we apply the snapshot and check mechanisms. 
In the event that a ransomware sample attempts to launch an attack, the system could be rolled back to the initial state by starting a snapshot.
This setup allows us to perform dynamic analysis of samples in a more secure manner. For RQ4, we deployed the prototype of \toolname on 3.32 million end users to detect ransomware over three months~(from March to June) to demonstrate its usefulness in a real industrial environment.

\subsubsection{Datasets} 
To ensure high-quality evaluation, we carefully curate two comprehensive datasets, one for training and testing, including both ransomware and benign samples, and the other for benchmarking performance comparison, containing only ransomware samples.

\noindent \textbf{Training \& Testing Dataset.} To assess the effectiveness of the training process, we curate a diverse dataset consisting of ransomware samples and benign files.
We collect 1,335 ransomware samples based on real cases from Bazaar\footnote{https://bazaar.abuse.ch} and VirusTotal\footnote{https://www.virustotal.com}, covering the period from January 2018 to December 2022.
These samples are used to form the training dataset.
To evaluate \toolname's generalization performance on previously unseen and unknown ransomware variants, we curate a distinct test dataset comprising 426 actively exploited ransomware samples from the wild. These samples, entirely separate from our training set, represent current threats that are still evading detection by all engines on VirusTotal. These elusive ransomware variants exemplify zero-day attacks, where adversaries exploit vulnerabilities before they become publicly known, making them particularly challenging to identify and mitigate.
For normal samples, we obtain 1,768 commonly used client-side and industry-specific applications (office, medical, and design software) and collect their behavioral characteristics through a sandbox virtual execution engine and operating system environment simulation.
We deliberately exclude benign samples such as zip, encryption, and backup software to avoid behaviors similar to ransomware.
The specific statistics are shown in \autoref{tab1}.

\begin{table}[!t]
	\centering	
	\caption{Training \& Testing dataset statistics.}
	\label{tab1}
     \resizebox{0.93\linewidth}{!}{
	\begin{tabular}{lr}
		\hline
		\textbf{Dataset Components} & \textbf{Quantity} \\ \hline
		Known ransomware samples & 1,335 \\
		Unknown ransomware samples & 426 \\
		Normal samples & 1,768 \\
		Known ransomware behavior data & 1,602,775 \\
		Unknown ransomware behavior data & 555,678 \\
		Normal behavior data & 1,885,490 \\ \hline
	\end{tabular}
 }
\end{table}

\noindent \textbf{Monthly Public Dataset.}
To conduct a comparative study for RQ2, we collect a monthly public dataset comprising real-world ransomware samples reported by users and from open-source websites. This dataset consists of 250 ransomware attacks targeting user endpoints, spanning a period from February to June 2023. These samples include some common ransomware families such as Phobos, LockBit, and Conti, which have been widely documented and observed in the wild. Additionally, the dataset incorporates twelve classes of emerging ransomware families such as PayMe100USD. These emerging families conceal their behavioral characteristics through process injection, making it challenging for security tools to differentiate between benign and malicious processes, and ultimately hindering their ability to detect these stealthy attacks. 

\subsubsection{Evaluation Metric}
To evaluate the performance, we use the following effectiveness metrics: True Positives~(TPs),
False Positives~(FPs), True Negatives~(TNs),
and False Negatives~(FNs). The specific metrics for evaluation are:

$\bullet$ True Positive Rate~(TPR) represents the percentage of ransomware attacks that are correctly detected and identified.
	
$\bullet$ False Positive Rate~(FPR) measures the percentage of benign files that are incorrectly classified as ransomware.

$\bullet$ True Negative Rate~(TNR) represents the percentage of benign files that are correctly identified as non-ransomware.

$\bullet$ False Negative Rate~(FNR), on the other hand, measures the percentage of ransomware attacks that are incorrectly classified as legitimate by the detection system.

\subsection{RQ1: Effectiveness}


\begin{table}[!t]
    \centering	
    \caption{Overall experimental performance.}
    \label{tab2}
    \resizebox{0.9\linewidth}{!}{
    \begin{tabular}{ll}
        \hline
        \textbf{Metric}               & \textbf{Result} \\ \hline
        Training time  & 10s    \\
        Model size     & 29.3KB \\
        Inference time & 30ms   \\
        Maximum response time  &  3s  \\
        Maximum file loss  &  1.21\%  \\ 
        TPs / TPR on training dataset    & 1,599,994 / 99.83\%     \\
        FNs / FNR on training dataset   & 2,781 / 0.17\%      \\
        TPs / TPR on test dataset  & 553,740 / 99.65\%       \\
        FNs / FNR on test dataset & 1,938 / 0.35\%      \\
        FP / FPR                  & 7 / $3.71e^{-6}$         \\ \hline   
    \end{tabular}
    }
\end{table}

\noindent \textbf{Real-time Response.} 
By enabling the protection of \toolname, we execute samples in a sandbox environment and collect behavioral data on suspicious processes that trigger monitoring points. The data collection utilizes a sliding window of 1 second, allowing for granular analysis of process behavior over time. As shown in \autoref{tab2}, \toolname demonstrates exceptional efficiency, with an average model inference time of merely 30ms. This rapid response ensures that any potential ransomware threat can be detected and isolated within up to 3 seconds after initial execution. Even when confronted with LockBit, currently the fastest ransomware~\cite{splunk2023speed}, \toolname can successfully prevent the vast majority of file encryption, limiting the overall file loss to a mere 1.21\%. Considering the average encryption speed reported by Splunk~\cite{splunk2023speed}, user file loss can even be minimized to a mere 0.12\%.

\noindent \textbf{Lightweight Framework.}
We employ a lightweight, funnel-type detection framework for \toolname, designed to strike a balance between effectiveness and efficiency. As shown in \autoref{tab2}, the training process for this framework is remarkably fast, taking only 10 seconds to complete. Moreover, the resulting model has a compact size of merely 29.3KB, making it highly portable and resource-efficient.
The low computational requirements and minimal model size of our funnel-type detection framework make \toolname particularly well-suited for deployment on endpoint devices, without imposing significant overhead or compromising system performance. 

\noindent \textbf{High Precision for Zero-Day Ransomware Detection.}
We leverage a multi-layered approach, employing various behavioral engines to filter and identify malicious threats efficiently. For known ransomware samples in the training dataset, \toolname boasts an impressive TPR of 99.83\%, showcasing its ability to identify and mitigate previously documented ransomware variants accurately.
Remarkably, even when faced with these unknown ransomware or novel variants in the unseen test dataset, \toolname maintains a high TPR of 99.65\%. This outstanding performance underscores the system's exceptional threat disposal capability, highlighting its ability to adapt and effectively identify emerging ransomware threats that have not been encountered before.

\noindent \textbf{Near-Zero False Positives.} 
Existing anti-virus software systems often generate a significant number of false ransomware alerts, which lack credibility and reliability. These false positives not only increase the workload of security analysts but also disrupt normal business operations by flagging legitimate software as potential threats.
In contrast, \toolname's ability to maintain a near-zero false positive rate is particularly impressive, especially when considering the vast amount of benign software behavior data it has to analyze. As shown in \autoref{tab2}, \toolname correctly classifies an astonishing 1,885,483 events of legitimate software, with only 7 FPs, resulting in an extremely low FPR of $3.71e^{-6}$.
By minimizing false positive alerts, \toolname significantly reduces the noise and unnecessary interruptions that often plague traditional anti-virus systems.

\begin{table*}[!t]
\def\arraystretch{1.35}
\centering
\caption{Comparison with SOTA academic solutions. Note that DFM represents decoy file monitor; RNM represents ransom note monitor; and FBD represents file behavior detection. \Circle ~represents that the tool does not support this feature; \RIGHTcircle ~represents that the tool supports this feature, but there is a gap when compared to the SOTA techniques; \CIRCLE ~denotes the SOTA techniques. }
\label{tab:academic}
\fontsize{8.5}{11}\selectfont
\begin{threeparttable}
\begin{tabular}{c||c||c||c||c||cc||ccc||c}
\hline
\multirow{2}{*}{\textbf{Solutions}} & \multirow{2}{*}{\textbf{\begin{tabular}[c]{@{}c@{}}Dataset\\ Size\end{tabular}}} & \multirow{2}{*}{\textbf{\begin{tabular}[c]{@{}c@{}}Realtime\\ Detection\end{tabular}}} & \multirow{2}{*}{\textbf{File-Loss}} & \multirow{2}{*}{\textbf{\begin{tabular}[c]{@{}c@{}}Runtime\\ Overhead\end{tabular}}} & \multicolumn{2}{c||}{\textbf{Effectiveness}}                           & \multicolumn{3}{c||}{\textbf{Approach}}                                                                                                     & \multirow{2}{*}{\textbf{\begin{tabular}[c]{@{}c@{}}Anti-\\ Bypass\end{tabular}}} \\ \cline{6-10}
                                    &                                                                                  &                                                                                        &                                     &                                                                                      & \multicolumn{1}{c|}{\textbf{FPR}}                   & \textbf{FNR}                      & \multicolumn{1}{c|}{\textbf{DFM}}                          & \multicolumn{1}{c|}{\textbf{RNM}}                    & \textbf{FBD}                    &                                                                                       \\ \hline
Unveil~\cite{DBLP:conf/uss/KharrazAMRK16}                              & 319                                                                              & \Circle                                                                  & /                                   & /                                                                                    & \multicolumn{1}{c|}{0.00\%} & $41\%^\clubsuit $ & \multicolumn{1}{c|}{\Circle}          & \multicolumn{1}{c|}{\Circle}          & \CIRCLE          & \Circle                                                                 \\
Redemption~\cite{DBLP:conf/raid/KharrazK17}                          & 677                                                                              & \Circle                                                                  & /                                   & 2.80\%\string~8.70\%                                                                          & \multicolumn{1}{c|}{1.00\%} & $69\%^\clubsuit $   & \multicolumn{1}{c|}{\Circle}          & \multicolumn{1}{c|}{\Circle}          & \CIRCLE          & \Circle                                                                 \\
ShieldFS~\cite{DBLP:conf/acsac/ContinellaGZPBZ16}                            & 305                                                                              & \Circle                                                                  & /                                   & 30\%\string~380\%                                                                                 & \multicolumn{1}{c|}{0.00\%}                   & 2.30\%                      & \multicolumn{1}{c|}{\Circle}          & \multicolumn{1}{c|}{\Circle}          & \CIRCLE          & \Circle                                                                 \\
RWGuard~\cite{DBLP:conf/raid/MehnazMB18}                             & 14                                                                               & \RIGHTcircle~(8.87s)                                                           & $2.79\%^\clubsuit $                             & 1.90\%                                                                               & \multicolumn{1}{c|}{0.10\%}                & 0.00\%                      & \multicolumn{1}{c|}{\CIRCLE}          & \multicolumn{1}{c|}{\Circle}          & \CIRCLE          & \RIGHTcircle                                                                 \\
RTrap~\cite{DBLP:journals/tifs/GanfureWCS23}                               & 1,106                                                                            & \RIGHTcircle~(5.35s\string~11.00s)                                                       & 0.17\%                              & /                                                                                    & \multicolumn{1}{c|}{$77.34\%^\star$}                     & /                        & \multicolumn{1}{c|}{\CIRCLE}          & \multicolumn{1}{c|}{\Circle}          & \Circle          & \CIRCLE                                                                 \\
\textbf{\toolname}                   & \textbf{1,761}                                                                   & \textbf{\CIRCLE~(\textless{}3.00s)}                                          & \textbf{\textless{}0.12\%\string~1.21\%}          & \textbf{0.60\%}                                                                       & \multicolumn{1}{c|}{\textbf{0.17\%}}       & \textbf{0.35\%}          & \multicolumn{1}{c|}{\textbf{\CIRCLE}} & \multicolumn{1}{c|}{\textbf{\CIRCLE}} & \textbf{\CIRCLE} & \textbf{\CIRCLE}                                                        \\ \hline
\end{tabular}
\begin{tablenotes}
  \item[$\clubsuit $]{These data marked with $\clubsuit $ denote results that have been reproduced by the latest studies, rather than the original reported results. The FNR of Unveil and Redemption have been reproduced by MarauderMap~\cite{hou2024empirical}; the file-loss of RWGuard has been reproduced by RTrap~\cite{DBLP:journals/tifs/GanfureWCS23}.} 
  \item[$\star$]{Due to the fact that RTrap is a solution solely based on DFM, the calculation method for its FPR is different from others. So we calculate the FPR of decoy files based on a continuous deployment in an industrial environment for 9 months.}
\end{tablenotes}
\end{threeparttable}
\end{table*}

\begin{table*}[!t]
\def\arraystretch{1.35}
\centering
\caption{Monthly performance test when compared with SOTA industrial solutions. }
\label{tab:industrial}
\fontsize{8.5}{11}\selectfont
\begin{threeparttable}
\begin{tabular}{c||c||c||ccccc}
\hline
\textbf{Month}            & \textbf{Technique}             & \textbf{Quantity} & \textbf{\toolname} & \textbf{Solution A} & \textbf{Solution B} & \textbf{Solution C} & \textbf{Solution D} \\ \hline
Feb-2023                  & Non-process Injection & 33~(6 families)                & \textbf{32~(96.97\%)}                        & 30~(90.91\%)         & 20~(60.61\%)        & 31~(93.94\%)   & 31~(93.94\%)       \\ \hline
Mar-2023                  & Non-process Injection & 49~(5 families)                & \textbf{48~(97.96\%)}                        & 47~(95.92\%)         & 29~(59.18\%)        & 45~(91.84\%)   & 43~(87.76\%)       \\ \hline
\multirow{2}{*}{Apr-2023} & Non-process Injection & 25~(10 families)                & \textbf{25~(100.00\%) }                      & \textbf{25~(100.00\%)}        & 18~(72.00\%)        & 20~(80.00\%)   & 22~(88.00\%)       \\
                          & Process Injection     & 31~(5 families)                & \textbf{31~(100.00\%)}                       & \textbf{31~(100.00\%)}        & 21~(67.74\%)        & 0~(0.00\%)     & 2~(6.45\%)         \\ \hline
\multirow{2}{*}{May-2023} & Non-process Injection & 36~(9 families)                & \textbf{36~(100.00\%)}                       & \textbf{36~(100.00\%)}        & \textbf{36~(100.00\%)}       & 33~(91.67\%)   & 35~(97.22\%)       \\
                          & Process Injection     & 12~(6 families)                & \textbf{12~(100.00\%)}                       & 10~(83.33\%)         & 6~(50.00\%)         & 4~(33.33\%)    & 4~(33.33\%)        \\ \hline
\multirow{2}{*}{Jun-2023} & Non-process Injection & 50~(5 families)                & \textbf{50~(100.00\%)}                       & 48~(96.00\%)        & 46~(92.00\%)        & 42~(84.00\%)   & 36~(72.00\%)       \\
                          & Process Injection     & 14~(6 families)                & \textbf{14~(100.00\%)}                       & 12~(85.71\%)         & 9~(64.29\%)         & 5~(35.71\%)    & 7~(50.00\%)        \\ \hline
\end{tabular}
\begin{tablenotes}
  \item[1]{To mitigate the impact of virus signature updates, we freeze the versions of all tested solutions as of February 2023.} 
\end{tablenotes}
\end{threeparttable}
\end{table*}

\subsection{RQ2: Comparison with SOTA Techniques}

To further demonstrate the superiority of \toolname, we conduct a comparative study against state-of-the-art~(SOTA) ransomware detection tools from both academia and industry.

\noindent \textbf{Comparison with Academic Solutions.} We compare \toolname with five SOTA academic solutions, i.e., Unveil~\cite{DBLP:conf/uss/KharrazAMRK16}, Redemption~\cite{DBLP:conf/raid/KharrazK17}, ShieldFS~\cite{DBLP:conf/acsac/ContinellaGZPBZ16},  RWGuard~\cite{DBLP:conf/raid/MehnazMB18}, and RTrap~\cite{DBLP:journals/tifs/GanfureWCS23}. The comparison results are shown in \autoref{tab:academic}. Note that as the SOTA research systems are not publicly available, we could not evaluate them in a real experimental environment. Therefore, we collect public evaluation data of the research systems from their publications and then make a comparison with \toolname. 
As shown in \autoref{tab:academic}, we compare various key performance metrics, including dataset size, real-time capability, file loss, runtime overhead, effectiveness, and capability to combat bypasses. Compared to solutions relying solely on file behavior monitoring, such as ShieldFS, \toolname significantly reduces the runtime overhead associated with comprehensive process behavior monitoring while offering superior effectiveness, particularly in terms of the FNR. Although Unveil and Redemption reported near-zero FNR in their original studies, the reproduced results from MarauderMap indicate their poor generalization capabilities on the latest ransomware dataset. In contrast to the latest advancements in bait-based ransomware detection, \toolname introduces a novel approach that combines ransomware note semantic detection and file behavior monitoring. This integrated approach effectively filters out false positives that may arise from bait triggers in solutions like RTrap, while also addressing the vulnerability of RWGuard to evasion techniques that exploit file access priority policies. Furthermore, \toolname excels in real-time detection capabilities, outperforming bait-based academic solutions in terms of responsiveness. 

\noindent \textbf{Comparison with Industry Solutions.} To further evaluate the effectiveness of \toolname, we conduct monthly performance tests.
By comparing the system detection rates for different months, we can track the effectiveness of our model over time and identify any potential weaknesses or areas for improvement. 
We select international cutting-edge endpoint security vendors Solution A; well-known Chinese domestic vendors Solution B, Solution C, and Solution D to compare the ransomware protection performance. 
We calculate the system detection rate according to the month of ransomware attack events.
The statistical results are shown in \autoref{tab:industrial}. Detailed test results of each family from February to June 2023 can be found in 
\crappendix{D}.
The results show that our model demonstrates competitive performance with Solution A, the world's leading anti-virus software, in blocking the majority of threats in non-process injection ransomware through scanning and detection. 
In process injection ransomware, our model achieves a high-risk virus detection rate of 100\% with high accuracy, by using fine-grained tracing of the parent process and its context, combined with the ransomware decoy protection engine, to terminate abnormal processes immediately.
Overall, our continuous test on monthly public datasets demonstrates the competitive performance of our model in detecting both known and unknown ransomware threats. 
By doing so, we can ensure that our defense systems remain up-to-date and effective in the face of evolving threats.

\subsection{RQ3: Ablation Study}
To gain a deeper understanding of the individual contributions of the key components in \toolname's ransomware detection pipeline, we conduct an ablation study. This systematic analysis involves selectively disabling or removing specific modules or techniques within the pipeline and evaluating the impact on overall performance. Specifically, we evaluate the following configurations:

\noindent \textbf{Variant\#1~(w/o DFM and RNM):} In this configuration, we disable the decoy file monitor and ransom note monitor module. In the absence of monitoring points, Variant\#1 have to monitor the file access behavior of all processes, resulting in significant overhead and resource consumption. The provided data in \autoref{tab:resource_utilization} highlights the stark contrast in resource utilization between scenarios with and without monitoring points. When logging all file behavior without monitoring points on a standard office computer (Win10, 2 CPU cores, 4GB RAM, 30GB HDD), the average CPU utilization soars to 6.70\%, with peak levels reaching a staggering 26.60\%. This substantial CPU usage can potentially impact the overall system performance and responsiveness, especially on resource-constrained office systems.

\begin{table}[!t]
\def\arraystretch{1.25}
\centering
\caption{Resource utilization comparison of Variant\#1.}
\begin{tabular}{ccc}
\toprule
\multicolumn{1}{c}{\textbf{Resource}} & \textbf{Variant\#1} & \textbf{\toolname} \\
\midrule
Average CPU Utilization & 6.70\% & 0.60\% \\
Peak CPU Utilization & 26.60\% & 6.20\% \\
Disk Consumption & 78MB & Negligible \\
Running Time & 30s & 30s \\
\bottomrule
\end{tabular}
\label{tab:resource_utilization}
\end{table}

\begin{figure}[!t]
    \centering
    \includegraphics[width=0.99\linewidth]{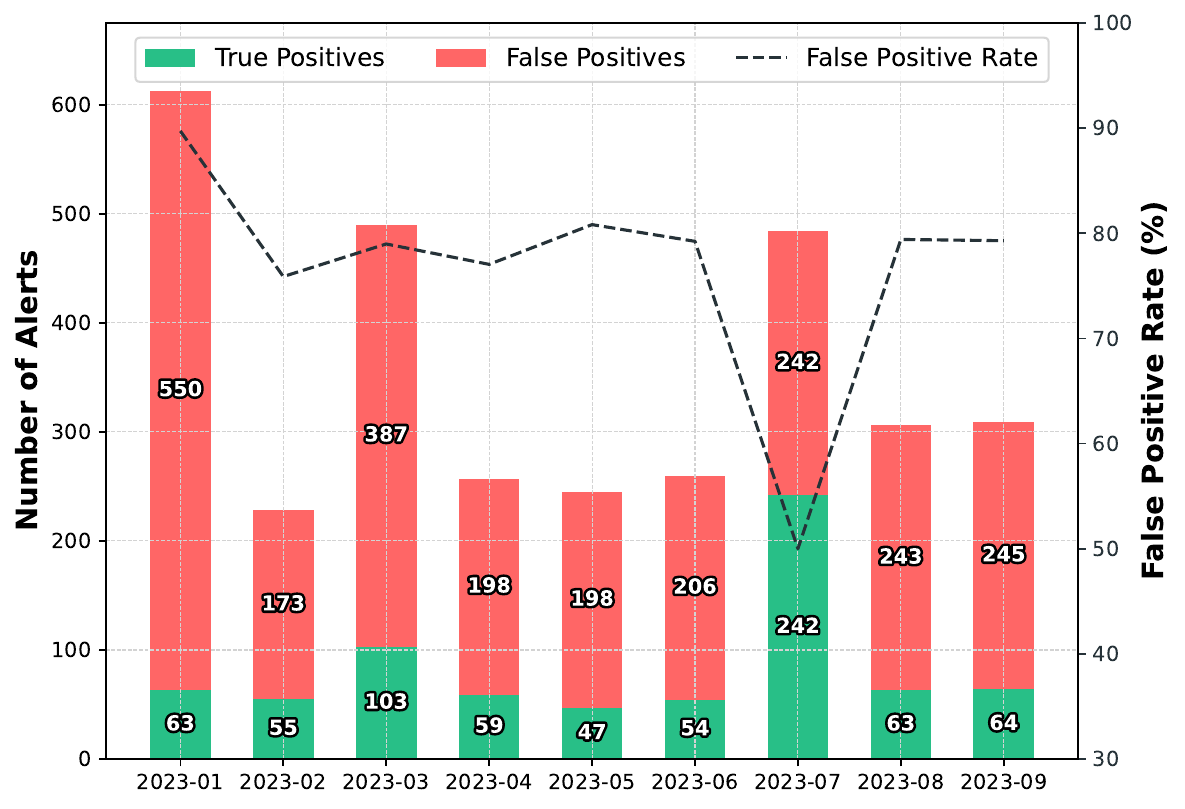}
    \vspace{-5pt}
    \caption{Performance of Variant\#2~ w/o FDB.}
    \label{fig:rq3-decoy}
\end{figure}

\noindent \textbf{Variant\#2~(w/o FBD):} In this variant, we remove the file behavior detector component, which tracks and analyzes changes to file systems, such as encryption, deletion, or modification of user data. As shown in \autoref{fig:rq3-decoy}, Variant\#2, which excludes the file behavior detector component, exhibits a significant increase in false positives across multiple months. 
The data provided indicates that from January 2023 to September 2023, the false positive rate for Variant\#2 remained consistently high, ranging from 75.88\% to 89.72\% for most months. With the FBD component in place, \toolname avoids 76.50\% of the inspection efforts required from security analysts, reducing the number of alerts that need manual review from 3,192 to 750~(per endpoint). This dramatic reduction in false positives not only enhances the accuracy of the detection system but also allows security personnel to focus on genuine threats rather than being inundated with false alarms.

\subsection{RQ4: Industrial Deployment}
By the time of this writing, \toolname has been successfully deployed on 3.32 million end users and has been running steadily for over a year. 
As shown in 
\crappendix{E}
, \toolname has successfully detected 61 ransomware attacks that occurred from March 2023 to April 2024, which are directed toward diverse industrial enterprises, encompassing manufacturing, technology, services, transportation, and additional domains.
Specifically, \toolname has defended against both well-known and distinctive variations of ransomware, including Mallox, Lockbit 3.0, and Tellyouthepass.
For further analysis, we conducted a manual forensic analysis of 27 ransomware attacks detected from March to June 2023.
These case studies showcase that \toolname can identify and defend a diverse range of intrusive techniques employed by ransomware, including the utilization of disguised files, the exploitation of vulnerabilities, and process injection.
Notably, it successfully thwarted 13 attacks leveraging n-day vulnerabilities documented in the CVE database, such as XVE-2023-3377 and QVD-2023-14179, as well as 5 high-risk attacks exploiting zero-day vulnerabilities.
This highlights the exceptional capability of \toolname for addressing unknown and sophisticated threats.
Overall, these results demonstrate that \toolname is highly effective in defending against real-world ransomware attacks on user endpoints and can provide reliable protection for various industries.
\section{Discussion}
\noindent \textbf{Potential Evasion Strategies}
In this study, \toolname relies on fixed monitoring points to trigger the subsequent behavioral engine. However, attackers may attempt to evade detection by the system. One potential evasion strategy is to avoid triggering monitoring points. Attackers may choose not to release ransom notes, modify decoy files, or perform file operations. 
In this particular situation, \toolname is unable to detect effectively.
We do recognize that ransomware detection is a dynamic adversarial process with no perfect solution. Indeed, this dynamic nature of the ransomware threat landscape motivates \toolname's modular, multi-layered design. While this paper focuses on decoy files and ransom notes, the actual engineering deployment could include additional strategies to counteract evasion tactics, such as monitoring magic number changes in file headers or file entropy changes before and after encryption. 
Although sophisticated evasion techniques may challenge individual monitoring methods, the multi-layered approach of \toolname significantly raises the bar for successful attacks. 

\noindent \textbf{Response Measures for Ransomware Incidents}
During the post-incident response phase, \toolname is capable of intercepting and blocking the ransomware, which involves terminating the encryption process and isolating the affected host from the network.
While this initial intervention provides immediate containment, it does not fully mitigate the underlying threat. The attacker may retain remote control over the compromised endpoint, creating a persistent access vector that enables potential repeated execution of the ransomware payload.
One possible solution is to terminate potential remote control services and enable two-factor authentication. 
This approach would significantly reduce the impact on user endpoints, particularly when compared to the current strategy of host isolation.
\section{Related Work}
Ransomware detection and response systems aim to identify and mitigate ransomware attacks on endpoints. These systems typically employ a combination of static analysis, behavior-based analysis, and decoy-based techniques.

\noindent \textbf{Static Analysis.}
The static detection method refers to analyzing static attributes such as byte sequences and opcodes without running the sample, extracting features from them, and training detection models. 
Medhat~\textit{et al.}~\cite{DBLP:conf/dasc/MedhatGA18} extract a series of static features (including file keywords, API functions, cryptographic signatures, file extensions, etc.).
Sheen~\textit{et al.}~\cite{DBLP:conf/icacci/SheenY18} summarize the frequency pattern of API file calls in malware.
Yamany~\textit{et al.}~\cite{yamany2022new} focus on raw ransomware binaries. They extract static feature sets, such as import address tables and strings, to identify common features and cluster similar samples.
Zhu~\textit{et al.}~\cite{DBLP:journals/compsec/ZhuJSWAC22} use more fine-grained code analysis to directly calculate the entropy features in ransomware binaries.
However, as technology evolves, attackers employ various methods such as compression, encryption, shelling, and obfuscation to prevent the extraction of static features. 
Aghakhani~\textit{et al.}~\cite{aghakhani2020packin} demonstrate that machine learning classifiers based on static analysis features are limited in their ability to generalize to unseen packers and are susceptible to adversarial examples, leading to false positives and potential mislabeling of ground-truth data.
While these static analysis approaches provide valuable insights, they are increasingly vulnerable to evasion techniques employed by modern ransomware. In contrast, \toolname adopts a dynamic approach that combines behavior-based and decoy-based techniques, offering a more robust and adaptive solution.

\noindent \textbf{Behavior-based Dynamic Analysis.}
Behavior-based dynamic analysis monitors system activities during program execution to detect ransomware. 
Scaife~\textit{et al.}~\cite{DBLP:conf/icdcs/ScaifeCTB16} track file activity patterns of read/write/delete operations to identify ransomware behavior. 
Kharraz~\textit{et al.}~\cite{DBLP:conf/raid/KharrazK17} propose Redemption, a system that monitors I/O request patterns on a per-process basis and maintains a transparent buffer for all storage I/O, allowing for immediate termination of suspicious processes and data restoration with minimal system modifications and performance overhead. 
Ma~\textit{et al.}~\cite{ma2023travelling} propose RansomTag, a hypervisor-based approach that employs fine-grained behavior analysis by introspecting OS-level context information, including file-related operations, process structures, and file objects, to accurately detect ransomware activities and enable precise data recovery. 
Similar approaches monitor call sequences of file access specifically for ransomware detection~\cite{DBLP:journals/corr/SgandurraMML16, DBLP:journals/istr/HamptonBZ18}. 
Berrueta~\textit{et al.}~\cite{DBLP:journals/eswa/BerruetaMMI22} combine file behavior monitoring with network communication analysis to detect ransomware activities, including command and control communications. 
However, these methods often suffer from high false positive rates, as some benign programs can exhibit similar file operation patterns to ransomware~\cite{zhou2023limits}. Existing approaches~\cite{DBLP:journals/compsec/Gomez-Hernandez18, DBLP:conf/raid/MehnazMB18, DBLP:journals/compsec/VorisSSHS19} are very sensitive to file operations, potentially misclassifying normal file-intensive programs as ransomware threats.
In contrast to these approaches, \toolname addresses the limitations of existing behavior-based methods by selectively filtering suspicious processes and performing in-depth behavioral analysis. This approach significantly reduces computational overhead and false positives, making it suitable for large-scale industrial applications.

\noindent \textbf{Decoy-based Dynamic Analysis.}
Decoy-based techniques deploy honeypot files or folders to detect ransomware activities. 
Moore~\cite{DBLP:conf/ccc2/Moore16} monitors file screens of honeypot folders to identify ransomware encryption attempts. 
Moussailebe~\textit{et al.}~\cite{DBLP:conf/IEEEares/MoussailebBPBCL18} deploy decoy folders in specific paths under the C drive to trap ransomware. 
Lee~\textit{et al.}~\cite{DBLP:conf/racs/LeeLH17} propose an efficient method for creating decoy files to detect ransomware, based on a source code-level analysis of ransomware behavior patterns, suggesting the strategic placement of two decoy files in the root directory to exploit common file traversal patterns observed in ransomware samples. 
G{\'{o}}mez{-}Hern{\'{a}}ndez~\textit{et al.}~\cite{DBLP:journals/compsec/Gomez-Hernandez18} introduce R-Locker, using named pipe decoy files, though its uniform nature might limit effectiveness.
Ganfure~\textit{et al.}~\cite{DBLP:journals/tifs/GanfureWCS23} develop RTrap, a machine learning-based framework that strategically places deceptive decoy files across directories to lure and detect ransomware, employing a lightweight decoy watcher for real-time monitoring and rapid automated response, achieving effective detection with minimal file loss. 
RWGuard~\cite{DBLP:conf/raid/MehnazMB18} combines decoy techniques with process and file system monitoring to provide real-time ransomware detection with high accuracy and low overhead. 
These approaches leverage the fact that ransomware typically attempts to encrypt all accessible files, including decoys. However, existing decoy-based methods often generate an overwhelming number of false positives in large-scale industrial clusters~\cite{DBLP:journals/tifs/GanfureWCS23,splunk2023speed,risk2021avoiding}, leading to intolerable disruptions and excessive inspection efforts. Moreover, they typically rely solely on decoy file interactions, potentially missing other ransomware indicators. In contrast, \toolname addresses these limitations with a multi-faceted monitoring approach. By combining the decoy-based monitor with the ransom note monitor, \toolname provides a more robust and reliable solution for ransomware detection in large-scale industrial applications.
\section{Conclusion}
In this paper, we propose \toolname, a lightweight and real-time ransomware detection system specifically designed for large-scale industrial environments. By innovatively prioritizing the analysis of suspicious processes through selective monitoring layers, \toolname effectively minimizes computational load and reduces alert fatigue without compromising detection capabilities. Furthermore, the rapid response and low CPU overhead provided by \toolname ensure its practicality for long-term, seamless deployment in industrial environments. The successful deployment across millions of endpoints and its proven track record in thwarting both known and zero-day ransomware attacks further attest to its robustness and reliability. In conclusion, \toolname offers promising prospects for safeguarding sensitive industrial environments against the growing menace of ransomware.
\section*{Acknowledgment}
This work was partly supported by the National Key R\&D Program of China~(2021YFB2701000), the Key R\&D Program of Hubei Province~(2023BAB017, 2023BAB079), the National NSF of China (grants No.62302181, 62072046),  the Knowledge Innovation Program of Wuhan-Basic Research, Huawei Research Fund, and HUSTCSE-FiberHome Joint Research Center for Network Security.

\newpage
\bibliographystyle{ACM-Reference-Format}
\balance
\bibliography{reference}


\begin{thebibliography}{43}


\ifx \showCODEN    \undefined \def \showCODEN     #1{\unskip}     \fi
\ifx \showDOI      \undefined \def \showDOI       #1{#1}\fi
\ifx \showISBNx    \undefined \def \showISBNx     #1{\unskip}     \fi
\ifx \showISBNxiii \undefined \def \showISBNxiii  #1{\unskip}     \fi
\ifx \showISSN     \undefined \def \showISSN      #1{\unskip}     \fi
\ifx \showLCCN     \undefined \def \showLCCN      #1{\unskip}     \fi
\ifx \shownote     \undefined \def \shownote      #1{#1}          \fi
\ifx \showarticletitle \undefined \def \showarticletitle #1{#1}   \fi
\ifx \showURL      \undefined \def \showURL       {\relax}        \fi
\providecommand\bibfield[2]{#2}
\providecommand\bibinfo[2]{#2}
\providecommand\natexlab[1]{#1}
\providecommand\showeprint[2][]{arXiv:#2}

\bibitem[Aghakhani et~al\mbox{.}({[n.\,d.]})]%
        {aghakhani2020packin}
\bibfield{author}{\bibinfo{person}{Hojjat Aghakhani}, \bibinfo{person}{Fabio Gritti}, \bibinfo{person}{Francesco Mecca}, \bibinfo{person}{Martina Lindorfer}, \bibinfo{person}{Stefano Ortolani}, \bibinfo{person}{Davide Balzarotti}, \bibinfo{person}{Giovanni Vigna}, {and} \bibinfo{person}{Christopher Kruegel}.} \bibinfo{year}{[n.\,d.]}\natexlab{}.
\newblock \showarticletitle{When Malware is Packin' Heat; Limits of Machine Learning Classifiers Based on Static Analysis Features}.
\newblock \bibinfo{journal}{\emph{Network and Distributed Systems Security (NDSS) Symposium 2020}} (\bibinfo{year}{[n.\,d.]}).
\newblock
\urldef\tempurl%
\url{https://doi.org/10.14722/ndss.2020.24310}
\showDOI{\tempurl}


\bibitem[Ahmed et~al\mbox{.}(2020)]%
        {DBLP:journals/jnca/AhmedKHAH20}
\bibfield{author}{\bibinfo{person}{Yahye~Abukar Ahmed}, \bibinfo{person}{Baris Ko{\c{c}}er}, \bibinfo{person}{Md.~Shamsul Huda}, \bibinfo{person}{Bander Ali~Saleh Al{-}rimy}, {and} \bibinfo{person}{Mohammad~Mehedi Hassan}.} \bibinfo{year}{2020}\natexlab{}.
\newblock \showarticletitle{A system call refinement-based enhanced Minimum Redundancy Maximum Relevance method for ransomware early detection}.
\newblock \bibinfo{journal}{\emph{J. Netw. Comput. Appl.}}  \bibinfo{volume}{167} (\bibinfo{year}{2020}), \bibinfo{pages}{102753}.
\newblock
\urldef\tempurl%
\url{https://doi.org/10.1016/j.jnca.2020.102753}
\showDOI{\tempurl}


\bibitem[Alder(2023)]%
        {steve20230day}
\bibfield{author}{\bibinfo{person}{Steve Alder}.} \bibinfo{year}{2023}\natexlab{}.
\newblock \bibinfo{title}{Ransomware Gangs Increasingly Exploiting 0Day and 1Day Vulnerabilities}.
\newblock \bibinfo{howpublished}{\url{https://www.hipaajournal.com/ransomware-gangs-increasingly-exploiting-0day-and-1day-vulnerabilities/}}.
\newblock


\bibitem[Baker(2023)]%
        {kurt2023fileless}
\bibfield{author}{\bibinfo{person}{Kurt Baker}.} \bibinfo{year}{2023}\natexlab{}.
\newblock \bibinfo{title}{Fileless Malware Explained}.
\newblock \bibinfo{howpublished}{\url{https://www.crowdstrike.com/cybersecurity-101/malware/fileless-malware/}}.
\newblock


\bibitem[Berrueta et~al\mbox{.}(2022)]%
        {DBLP:journals/eswa/BerruetaMMI22}
\bibfield{author}{\bibinfo{person}{Eduardo Berrueta}, \bibinfo{person}{Daniel Morat{\'{o}}}, \bibinfo{person}{Eduardo Maga{\~{n}}a}, {and} \bibinfo{person}{Mikel Izal}.} \bibinfo{year}{2022}\natexlab{}.
\newblock \showarticletitle{Crypto-ransomware detection using machine learning models in file-sharing network scenarios with encrypted traffic}.
\newblock \bibinfo{journal}{\emph{Expert Syst. Appl.}}  \bibinfo{volume}{209} (\bibinfo{year}{2022}), \bibinfo{pages}{118299}.
\newblock
\urldef\tempurl%
\url{https://doi.org/10.1016/j.eswa.2022.118299}
\showDOI{\tempurl}


\bibitem[Breiman et~al\mbox{.}(1984)]%
        {DBLP:books/wa/BreimanFOS84}
\bibfield{author}{\bibinfo{person}{Leo Breiman}, \bibinfo{person}{J.~H. Friedman}, \bibinfo{person}{Richard~A. Olshen}, {and} \bibinfo{person}{C.~J. Stone}.} \bibinfo{year}{1984}\natexlab{}.
\newblock \bibinfo{booktitle}{\emph{Classification and Regression Trees}}.
\newblock \bibinfo{publisher}{Wadsworth}.
\newblock
\showISBNx{0-534-98053-8}


\bibitem[Cen et~al\mbox{.}(2024)]%
        {cen2024ransomware}
\bibfield{author}{\bibinfo{person}{Mingcan Cen}, \bibinfo{person}{Frank Jiang}, \bibinfo{person}{Xingsheng Qin}, \bibinfo{person}{Qinghong Jiang}, {and} \bibinfo{person}{Robin Doss}.} \bibinfo{year}{2024}\natexlab{}.
\newblock \showarticletitle{Ransomware early detection: A survey}.
\newblock \bibinfo{journal}{\emph{Computer Networks}}  \bibinfo{volume}{239} (\bibinfo{year}{2024}), \bibinfo{pages}{110138}.
\newblock


\bibitem[Chen and Guestrin(2016)]%
        {chen2016xgboost}
\bibfield{author}{\bibinfo{person}{Tianqi Chen} {and} \bibinfo{person}{Carlos Guestrin}.} \bibinfo{year}{2016}\natexlab{}.
\newblock \showarticletitle{XGBoost: A Scalable Tree Boosting System}. In \bibinfo{booktitle}{\emph{Proceedings of the 22nd ACM SIGKDD International Conference on Knowledge Discovery and Data Mining}} (San Francisco, California, USA) \emph{(\bibinfo{series}{KDD '16})}. \bibinfo{publisher}{Association for Computing Machinery}, \bibinfo{address}{New York, NY, USA}, \bibinfo{pages}{785–794}.
\newblock
\showISBNx{9781450342322}
\urldef\tempurl%
\url{https://doi.org/10.1145/2939672.2939785}
\showDOI{\tempurl}


\bibitem[Continella et~al\mbox{.}(2016)]%
        {DBLP:conf/acsac/ContinellaGZPBZ16}
\bibfield{author}{\bibinfo{person}{Andrea Continella}, \bibinfo{person}{Alessandro Guagnelli}, \bibinfo{person}{Giovanni Zingaro}, \bibinfo{person}{Giulio~De Pasquale}, \bibinfo{person}{Alessandro Barenghi}, \bibinfo{person}{Stefano Zanero}, {and} \bibinfo{person}{Federico Maggi}.} \bibinfo{year}{2016}\natexlab{}.
\newblock \showarticletitle{ShieldFS: a self-healing, ransomware-aware filesystem}. In \bibinfo{booktitle}{\emph{Proceedings of the 32nd Annual Conference on Computer Security Applications, {ACSAC} 2016, Los Angeles, CA, USA, December 5-9, 2016}}, \bibfield{editor}{\bibinfo{person}{Stephen Schwab}, \bibinfo{person}{William~K. Robertson}, {and} \bibinfo{person}{Davide Balzarotti}} (Eds.). \bibinfo{publisher}{{ACM}}, \bibinfo{pages}{336--347}.
\newblock
\urldef\tempurl%
\url{http://dl.acm.org/citation.cfm?id=2991110}
\showURL{%
\tempurl}


\bibitem[Davis(2022)]%
        {splunk2023speed}
\bibfield{author}{\bibinfo{person}{Shannon Davis}.} \bibinfo{year}{2022}\natexlab{}.
\newblock \bibinfo{title}{An Empirically Comparative Analysis of Ransomware Binaries}.
\newblock \bibinfo{howpublished}{\url{https://www.splunk.com/content/dam/splunk2/en_us/gated/white-paper/an-empirically-comparative-analysis-of-ransomware-binaries.pdf}}.
\newblock


\bibitem[Denham and Thompson(2023)]%
        {DBLP:conf/cns/DenhamT23}
\bibfield{author}{\bibinfo{person}{Byron Denham} {and} \bibinfo{person}{Dale~R. Thompson}.} \bibinfo{year}{2023}\natexlab{}.
\newblock \showarticletitle{Analysis of Decoy Strategies for Detecting Ransomware}. In \bibinfo{booktitle}{\emph{{IEEE} Conference on Communications and Network Security, {CNS} 2023, Orlando, FL, USA, October 2-5, 2023}}. \bibinfo{publisher}{{IEEE}}, \bibinfo{pages}{1--6}.
\newblock
\urldef\tempurl%
\url{https://doi.org/10.1109/CNS59707.2023.10288691}
\showDOI{\tempurl}


\bibitem[Fernando and Komninos(2022)]%
        {DBLP:journals/compsec/FernandoK22}
\bibfield{author}{\bibinfo{person}{Damien~Warren Fernando} {and} \bibinfo{person}{Nikos Komninos}.} \bibinfo{year}{2022}\natexlab{}.
\newblock \showarticletitle{FeSA: Feature selection architecture for ransomware detection under concept drift}.
\newblock \bibinfo{journal}{\emph{Comput. Secur.}}  \bibinfo{volume}{116} (\bibinfo{year}{2022}), \bibinfo{pages}{102659}.
\newblock
\urldef\tempurl%
\url{https://doi.org/10.1016/j.cose.2022.102659}
\showDOI{\tempurl}


\bibitem[Friedman(2001)]%
        {Friedman2001GreedyFA}
\bibfield{author}{\bibinfo{person}{Jerome~H. Friedman}.} \bibinfo{year}{2001}\natexlab{}.
\newblock \showarticletitle{Greedy function approximation: A gradient boosting machine.}
\newblock \bibinfo{journal}{\emph{Annals of Statistics}}  \bibinfo{volume}{29} (\bibinfo{year}{2001}), \bibinfo{pages}{1189--1232}.
\newblock


\bibitem[Ganfure et~al\mbox{.}(2023)]%
        {DBLP:journals/tifs/GanfureWCS23}
\bibfield{author}{\bibinfo{person}{Gaddisa~Olani Ganfure}, \bibinfo{person}{Chun{-}Feng Wu}, \bibinfo{person}{Yuan{-}Hao Chang}, {and} \bibinfo{person}{Wei{-}Kuan Shih}.} \bibinfo{year}{2023}\natexlab{}.
\newblock \showarticletitle{RTrap: Trapping and Containing Ransomware With Machine Learning}.
\newblock \bibinfo{journal}{\emph{{IEEE} Trans. Inf. Forensics Secur.}}  \bibinfo{volume}{18} (\bibinfo{year}{2023}), \bibinfo{pages}{1433--1448}.
\newblock
\urldef\tempurl%
\url{https://doi.org/10.1109/TIFS.2023.3240025}
\showDOI{\tempurl}


\bibitem[G{\'{o}}mez{-}Hern{\'{a}}ndez et~al\mbox{.}(2018)]%
        {DBLP:journals/compsec/Gomez-Hernandez18}
\bibfield{author}{\bibinfo{person}{Jos{\'{e}}~Antonio G{\'{o}}mez{-}Hern{\'{a}}ndez}, \bibinfo{person}{L. {\'{A}}lvarez{-}Gonz{\'{a}}lez}, {and} \bibinfo{person}{Pedro Garc{\'{\i}}a{-}Teodoro}.} \bibinfo{year}{2018}\natexlab{}.
\newblock \showarticletitle{R-Locker: Thwarting ransomware action through a honeyfile-based approach}.
\newblock \bibinfo{journal}{\emph{Comput. Secur.}}  \bibinfo{volume}{73} (\bibinfo{year}{2018}), \bibinfo{pages}{389--398}.
\newblock
\urldef\tempurl%
\url{https://doi.org/10.1016/j.cose.2017.11.019}
\showDOI{\tempurl}


\bibitem[Hampton et~al\mbox{.}(2018)]%
        {DBLP:journals/istr/HamptonBZ18}
\bibfield{author}{\bibinfo{person}{Nikolai Hampton}, \bibinfo{person}{Zubair~A. Baig}, {and} \bibinfo{person}{Sherali Zeadally}.} \bibinfo{year}{2018}\natexlab{}.
\newblock \showarticletitle{Ransomware behavioural analysis on windows platforms}.
\newblock \bibinfo{journal}{\emph{J. Inf. Secur. Appl.}}  \bibinfo{volume}{40} (\bibinfo{year}{2018}), \bibinfo{pages}{44--51}.
\newblock
\urldef\tempurl%
\url{https://doi.org/10.1016/j.jisa.2018.02.008}
\showDOI{\tempurl}


\bibitem[Hou et~al\mbox{.}(2024)]%
        {hou2024empirical}
\bibfield{author}{\bibinfo{person}{Yiwei Hou}, \bibinfo{person}{Lihua Guo}, \bibinfo{person}{Chijin Zhou}, \bibinfo{person}{Yiwen Xu}, \bibinfo{person}{Zijing Yin}, \bibinfo{person}{Shanshan Li}, \bibinfo{person}{Chengnian Sun}, {and} \bibinfo{person}{Yu Jiang}.} \bibinfo{year}{2024}\natexlab{}.
\newblock \showarticletitle{An Empirical Study of Data Disruption by Ransomware Attacks}. In \bibinfo{booktitle}{\emph{Proceedings of the 46th IEEE/ACM International Conference on Software Engineering (ICSE'24)}} (Lisbon, Portugal). ACM.
\newblock
\urldef\tempurl%
\url{https://doi.org/10.1145/3597503.3639090}
\showDOI{\tempurl}


\bibitem[Kharraz et~al\mbox{.}(2016)]%
        {DBLP:conf/uss/KharrazAMRK16}
\bibfield{author}{\bibinfo{person}{Amin Kharraz}, \bibinfo{person}{Sajjad Arshad}, \bibinfo{person}{Collin Mulliner}, \bibinfo{person}{William~K. Robertson}, {and} \bibinfo{person}{Engin Kirda}.} \bibinfo{year}{2016}\natexlab{}.
\newblock \showarticletitle{{UNVEIL:} {A} Large-Scale, Automated Approach to Detecting Ransomware}. In \bibinfo{booktitle}{\emph{25th {USENIX} Security Symposium, {USENIX} Security 16, Austin, TX, USA, August 10-12, 2016}}, \bibfield{editor}{\bibinfo{person}{Thorsten Holz} {and} \bibinfo{person}{Stefan Savage}} (Eds.). \bibinfo{publisher}{{USENIX} Association}, \bibinfo{pages}{757--772}.
\newblock
\urldef\tempurl%
\url{https://www.usenix.org/conference/usenixsecurity16/technical-sessions/presentation/kharaz}
\showURL{%
\tempurl}


\bibitem[Kharraz and Kirda(2017)]%
        {DBLP:conf/raid/KharrazK17}
\bibfield{author}{\bibinfo{person}{Amin Kharraz} {and} \bibinfo{person}{Engin Kirda}.} \bibinfo{year}{2017}\natexlab{}.
\newblock \showarticletitle{Redemption: Real-Time Protection Against Ransomware at End-Hosts}. In \bibinfo{booktitle}{\emph{Research in Attacks, Intrusions, and Defenses - 20th International Symposium, {RAID} 2017, Atlanta, GA, USA, September 18-20, 2017, Proceedings}} \emph{(\bibinfo{series}{Lecture Notes in Computer Science}, Vol.~\bibinfo{volume}{10453})}, \bibfield{editor}{\bibinfo{person}{Marc Dacier}, \bibinfo{person}{Michael~D. Bailey}, \bibinfo{person}{Michalis Polychronakis}, {and} \bibinfo{person}{Manos Antonakakis}} (Eds.). \bibinfo{publisher}{Springer}, \bibinfo{pages}{98--119}.
\newblock
\urldef\tempurl%
\url{https://doi.org/10.1007/978-3-319-66332-6\_5}
\showDOI{\tempurl}


\bibitem[Kharraz et~al\mbox{.}(2015)]%
        {DBLP:conf/dimva/KharrazRBBK15}
\bibfield{author}{\bibinfo{person}{Amin Kharraz}, \bibinfo{person}{William~K. Robertson}, \bibinfo{person}{Davide Balzarotti}, \bibinfo{person}{Leyla Bilge}, {and} \bibinfo{person}{Engin Kirda}.} \bibinfo{year}{2015}\natexlab{}.
\newblock \showarticletitle{Cutting the Gordian Knot: {A} Look Under the Hood of Ransomware Attacks}. In \bibinfo{booktitle}{\emph{Detection of Intrusions and Malware, and Vulnerability Assessment - 12th International Conference, {DIMVA} 2015, Milan, Italy, July 9-10, 2015, Proceedings}} \emph{(\bibinfo{series}{Lecture Notes in Computer Science}, Vol.~\bibinfo{volume}{9148})}, \bibfield{editor}{\bibinfo{person}{Magnus Almgren}, \bibinfo{person}{Vincenzo Gulisano}, {and} \bibinfo{person}{Federico Maggi}} (Eds.). \bibinfo{publisher}{Springer}, \bibinfo{pages}{3--24}.
\newblock
\urldef\tempurl%
\url{https://doi.org/10.1007/978-3-319-20550-2\_1}
\showDOI{\tempurl}


\bibitem[Lee et~al\mbox{.}(2017)]%
        {DBLP:conf/racs/LeeLH17}
\bibfield{author}{\bibinfo{person}{Jeonghwan Lee}, \bibinfo{person}{Jinwoo Lee}, {and} \bibinfo{person}{Jiman Hong}.} \bibinfo{year}{2017}\natexlab{}.
\newblock \showarticletitle{How to Make Efficient Decoy Files for Ransomware Detection?}. In \bibinfo{booktitle}{\emph{Proceedings of the International Conference on Research in Adaptive and Convergent Systems, {RACS} 2017, Krakow, Poland, September 20-23, 2017}}. \bibinfo{publisher}{{ACM}}, \bibinfo{pages}{208--212}.
\newblock
\urldef\tempurl%
\url{https://doi.org/10.1145/3129676.3129713}
\showDOI{\tempurl}


\bibitem[Ma et~al\mbox{.}(2023)]%
        {ma2023travelling}
\bibfield{author}{\bibinfo{person}{Boyang Ma}, \bibinfo{person}{Yilin Yang}, \bibinfo{person}{Jinku Li}, \bibinfo{person}{Fengwei Zhang}, \bibinfo{person}{Wenbo Shen}, \bibinfo{person}{Yajin Zhou}, {and} \bibinfo{person}{Jianfeng Ma}.} \bibinfo{year}{2023}\natexlab{}.
\newblock \showarticletitle{Travelling the Hypervisor and SSD: A Tag-Based Approach Against Crypto Ransomware with Fine-Grained Data Recovery}. In \bibinfo{booktitle}{\emph{Proceedings of the 2023 ACM SIGSAC Conference on Computer and Communications Security}} (Copenhagen, Denmark) \emph{(\bibinfo{series}{CCS '23})}. \bibinfo{publisher}{Association for Computing Machinery}, \bibinfo{address}{New York, NY, USA}, \bibinfo{pages}{341–355}.
\newblock
\showISBNx{9798400700507}
\urldef\tempurl%
\url{https://doi.org/10.1145/3576915.3616665}
\showDOI{\tempurl}


\bibitem[Masum et~al\mbox{.}(2022)]%
        {DBLP:conf/ccwc/MasumFSQ0A22}
\bibfield{author}{\bibinfo{person}{Mohammad Masum}, \bibinfo{person}{Md. Jobair~Hossain Faruk}, \bibinfo{person}{Hossain Shahriar}, \bibinfo{person}{Kai Qian}, \bibinfo{person}{Dan Lo}, {and} \bibinfo{person}{Muhaiminul~Islam Adnan}.} \bibinfo{year}{2022}\natexlab{}.
\newblock \showarticletitle{Ransomware Classification and Detection With Machine Learning Algorithms}. In \bibinfo{booktitle}{\emph{12th {IEEE} Annual Computing and Communication Workshop and Conference, {CCWC} 2022, Las Vegas, NV, USA, January 26-29, 2022}}. \bibinfo{publisher}{{IEEE}}, \bibinfo{pages}{316--322}.
\newblock
\urldef\tempurl%
\url{https://doi.org/10.1109/CCWC54503.2022.9720869}
\showDOI{\tempurl}


\bibitem[Medhat et~al\mbox{.}(2018)]%
        {DBLP:conf/dasc/MedhatGA18}
\bibfield{author}{\bibinfo{person}{May Medhat}, \bibinfo{person}{Samir Gaber}, {and} \bibinfo{person}{Nashwa Abdelbaki}.} \bibinfo{year}{2018}\natexlab{}.
\newblock \showarticletitle{A New Static-Based Framework for Ransomware Detection}. In \bibinfo{booktitle}{\emph{2018 {IEEE} 16th Intl Conf on Dependable, Autonomic and Secure Computing, 16th Intl Conf on Pervasive Intelligence and Computing, 4th Intl Conf on Big Data Intelligence and Computing and Cyber Science and Technology Congress, DASC/PiCom/DataCom/CyberSciTech 2018, Athens, Greece, August 12-15, 2018}}. \bibinfo{publisher}{{IEEE} Computer Society}, \bibinfo{pages}{710--715}.
\newblock
\urldef\tempurl%
\url{https://doi.org/10.1109/DASC/PiCom/DataCom/CyberSciTec.2018.00124}
\showDOI{\tempurl}


\bibitem[Mehnaz et~al\mbox{.}(2018)]%
        {DBLP:conf/raid/MehnazMB18}
\bibfield{author}{\bibinfo{person}{Shagufta Mehnaz}, \bibinfo{person}{Anand Mudgerikar}, {and} \bibinfo{person}{Elisa Bertino}.} \bibinfo{year}{2018}\natexlab{}.
\newblock \showarticletitle{RWGuard: {A} Real-Time Detection System Against Cryptographic Ransomware}. In \bibinfo{booktitle}{\emph{Research in Attacks, Intrusions, and Defenses - 21st International Symposium, {RAID} 2018, Heraklion, Crete, Greece, September 10-12, 2018, Proceedings}} \emph{(\bibinfo{series}{Lecture Notes in Computer Science}, Vol.~\bibinfo{volume}{11050})}, \bibfield{editor}{\bibinfo{person}{Michael Bailey}, \bibinfo{person}{Thorsten Holz}, \bibinfo{person}{Manolis Stamatogiannakis}, {and} \bibinfo{person}{Sotiris Ioannidis}} (Eds.). \bibinfo{publisher}{Springer}, \bibinfo{pages}{114--136}.
\newblock
\urldef\tempurl%
\url{https://doi.org/10.1007/978-3-030-00470-5\_6}
\showDOI{\tempurl}


\bibitem[Moore(2016)]%
        {DBLP:conf/ccc2/Moore16}
\bibfield{author}{\bibinfo{person}{Chris Moore}.} \bibinfo{year}{2016}\natexlab{}.
\newblock \showarticletitle{Detecting Ransomware with Honeypot Techniques}. In \bibinfo{booktitle}{\emph{Cybersecurity and Cyberforensics Conference, {CCC} 2016, Amman, Jordan, August 2-4, 2016}}. \bibinfo{publisher}{{IEEE}}, \bibinfo{pages}{77--81}.
\newblock
\urldef\tempurl%
\url{https://doi.org/10.1109/CCC.2016.14}
\showDOI{\tempurl}


\bibitem[Moussaileb et~al\mbox{.}(2018)]%
        {DBLP:conf/IEEEares/MoussailebBPBCL18}
\bibfield{author}{\bibinfo{person}{Routa Moussaileb}, \bibinfo{person}{Benjamin Bouget}, \bibinfo{person}{Aur{\'{e}}lien Palisse}, \bibinfo{person}{H{\'{e}}l{\`{e}}ne~Le Bouder}, \bibinfo{person}{Nora Cuppens}, {and} \bibinfo{person}{Jean{-}Louis Lanet}.} \bibinfo{year}{2018}\natexlab{}.
\newblock \showarticletitle{Ransomware's Early Mitigation Mechanisms}. In \bibinfo{booktitle}{\emph{Proceedings of the 13th International Conference on Availability, Reliability and Security, {ARES} 2018, Hamburg, Germany, August 27-30, 2018}}, \bibfield{editor}{\bibinfo{person}{Sebastian Doerr}, \bibinfo{person}{Mathias Fischer}, \bibinfo{person}{Sebastian Schrittwieser}, {and} \bibinfo{person}{Dominik Herrmann}} (Eds.). \bibinfo{publisher}{{ACM}}, \bibinfo{pages}{2:1--2:10}.
\newblock
\urldef\tempurl%
\url{https://doi.org/10.1145/3230833.3234691}
\showDOI{\tempurl}


\bibitem[Odusanya(2020)]%
        {mosimilolu2020injection}
\bibfield{author}{\bibinfo{person}{Mosimilolu Odusanya}.} \bibinfo{year}{2020}\natexlab{}.
\newblock \bibinfo{title}{MITRE ATT\&CK spotlight: Process injection}.
\newblock \bibinfo{howpublished}{\url{https://www.infosecinstitute.com/resources/mitre-attck/mitre-attck-spotlight-process-injection/}}.
\newblock


\bibitem[Oz et~al\mbox{.}(2022)]%
        {DBLP:journals/csur/OzALU22}
\bibfield{author}{\bibinfo{person}{Harun Oz}, \bibinfo{person}{Ahmet Aris}, \bibinfo{person}{Albert Levi}, {and} \bibinfo{person}{A.~Selcuk Uluagac}.} \bibinfo{year}{2022}\natexlab{}.
\newblock \showarticletitle{A Survey on Ransomware: Evolution, Taxonomy, and Defense Solutions}.
\newblock \bibinfo{journal}{\emph{{ACM} Comput. Surv.}} \bibinfo{volume}{54}, \bibinfo{number}{11s} (\bibinfo{year}{2022}), \bibinfo{pages}{238:1--238:37}.
\newblock
\urldef\tempurl%
\url{https://doi.org/10.1145/3514229}
\showDOI{\tempurl}


\bibitem[Razaulla et~al\mbox{.}(2023)]%
        {DBLP:journals/access/RazaullaFMGMFA23}
\bibfield{author}{\bibinfo{person}{Salwa Razaulla}, \bibinfo{person}{Claude Fachkha}, \bibinfo{person}{Christine Markarian}, \bibinfo{person}{Amjad Gawanmeh}, \bibinfo{person}{Wathiq Mansoor}, \bibinfo{person}{Benjamin C.~M. Fung}, {and} \bibinfo{person}{Chadi Assi}.} \bibinfo{year}{2023}\natexlab{}.
\newblock \showarticletitle{The Age of Ransomware: {A} Survey on the Evolution, Taxonomy, and Research Directions}.
\newblock \bibinfo{journal}{\emph{{IEEE} Access}}  \bibinfo{volume}{11} (\bibinfo{year}{2023}), \bibinfo{pages}{40698--40723}.
\newblock
\urldef\tempurl%
\url{https://doi.org/10.1109/ACCESS.2023.3268535}
\showDOI{\tempurl}


\bibitem[RiskOptics(2021)]%
        {risk2021avoiding}
\bibfield{author}{\bibinfo{person}{RiskOptics}.} \bibinfo{year}{2021}\natexlab{}.
\newblock \bibinfo{title}{Avoiding Cyber Security False Positives}.
\newblock \bibinfo{howpublished}{\url{https://reciprocity.com/blog/avoiding-cyber-security-false-positives/}}.
\newblock


\bibitem[Rosen et~al\mbox{.}(2023)]%
        {nvidia2023ransomware}
\bibfield{author}{\bibinfo{person}{Nir Rosen}, \bibinfo{person}{Haim Elisha}, \bibinfo{person}{Ahmad Saleh}, \bibinfo{person}{Vadim Gechman}, {and} \bibinfo{person}{Sharon Mashhadi}.} \bibinfo{year}{2023}\natexlab{}.
\newblock \bibinfo{title}{Supercharge Ransomware Detection with AI-Enhanced Cybersecurity Solutions}.
\newblock \bibinfo{howpublished}{\url{https://developer.nvidia.com/blog/supercharge-ransomware-detection-with-ai-enhanced-cybersecurity-solutions/}}.
\newblock


\bibitem[Scaife et~al\mbox{.}(2016)]%
        {DBLP:conf/icdcs/ScaifeCTB16}
\bibfield{author}{\bibinfo{person}{Nolen Scaife}, \bibinfo{person}{Henry Carter}, \bibinfo{person}{Patrick Traynor}, {and} \bibinfo{person}{Kevin R.~B. Butler}.} \bibinfo{year}{2016}\natexlab{}.
\newblock \showarticletitle{CryptoLock (and Drop It): Stopping Ransomware Attacks on User Data}. In \bibinfo{booktitle}{\emph{36th {IEEE} International Conference on Distributed Computing Systems, {ICDCS} 2016, Nara, Japan, June 27-30, 2016}}. \bibinfo{publisher}{{IEEE} Computer Society}, \bibinfo{pages}{303--312}.
\newblock
\urldef\tempurl%
\url{https://doi.org/10.1109/ICDCS.2016.46}
\showDOI{\tempurl}


\bibitem[Sgandurra et~al\mbox{.}(2016)]%
        {DBLP:journals/corr/SgandurraMML16}
\bibfield{author}{\bibinfo{person}{Daniele Sgandurra}, \bibinfo{person}{Luis Mu{\~{n}}oz{-}Gonz{\'{a}}lez}, \bibinfo{person}{Rabih Mohsen}, {and} \bibinfo{person}{Emil~C. Lupu}.} \bibinfo{year}{2016}\natexlab{}.
\newblock \showarticletitle{Automated Dynamic Analysis of Ransomware: Benefits, Limitations and use for Detection}.
\newblock \bibinfo{journal}{\emph{CoRR}}  \bibinfo{volume}{abs/1609.03020} (\bibinfo{year}{2016}).
\newblock
\showeprint[arXiv]{1609.03020}
\urldef\tempurl%
\url{http://arxiv.org/abs/1609.03020}
\showURL{%
\tempurl}


\bibitem[Sheen and Yadav(2018)]%
        {DBLP:conf/icacci/SheenY18}
\bibfield{author}{\bibinfo{person}{Shina Sheen} {and} \bibinfo{person}{Ashwitha Yadav}.} \bibinfo{year}{2018}\natexlab{}.
\newblock \showarticletitle{Ransomware detection by mining {API} call usage}. In \bibinfo{booktitle}{\emph{2018 International Conference on Advances in Computing, Communications and Informatics, {ICACCI} 2018, Bangalore, India, September 19-22, 2018}}. \bibinfo{publisher}{{IEEE}}, \bibinfo{pages}{983--987}.
\newblock
\urldef\tempurl%
\url{https://doi.org/10.1109/ICACCI.2018.8554938}
\showDOI{\tempurl}


\bibitem[Sophos(2023)]%
        {sophos2023ransomware}
\bibfield{author}{\bibinfo{person}{Sophos}.} \bibinfo{year}{2023}\natexlab{}.
\newblock \bibinfo{title}{The State of Ransomware 2023}.
\newblock \bibinfo{howpublished}{\url{https://assets.sophos.com/X24WTUEQ/at/c949g7693gsnjh9rb9gr8/sophos-state-of-ransomware-2023-wp.pdf}}.
\newblock


\bibitem[Steinberg et~al\mbox{.}(2002)]%
        {DBLP:conf/ausdm/SteinbergGC02}
\bibfield{author}{\bibinfo{person}{Dan Steinberg}, \bibinfo{person}{Mikhail Golovnya}, {and} \bibinfo{person}{Nicholas~Scott Cardell}.} \bibinfo{year}{2002}\natexlab{}.
\newblock \showarticletitle{Stochastic Gradient Boosting: An Introduction to TreeNet{\texttrademark}}. In \bibinfo{booktitle}{\emph{The 15th Australian Joint Conference on Artificial Intelligence 2002, Proceedings Australasian Data Mining Workshop, Canberra, Australia, 3rd December 2002}}, \bibfield{editor}{\bibinfo{person}{Simeon~J. Simoff}, \bibinfo{person}{Graham~J. Williams}, {and} \bibinfo{person}{Markus Hegland}} (Eds.). \bibinfo{publisher}{University of Technology Sydney, Australia}, \bibinfo{pages}{1--12}.
\newblock


\bibitem[Team(2024)]%
        {chainalysiss2024ransomware}
\bibfield{author}{\bibinfo{person}{Chainalysis Team}.} \bibinfo{year}{2024}\natexlab{}.
\newblock \bibinfo{title}{Ransomware Payments Exceed 1 Billion in 2023, Hitting Record High After 2022 Decline}.
\newblock \bibinfo{howpublished}{\url{https://www.chainalysis.com/blog/ransomware-2024/}}.
\newblock


\bibitem[Urooj et~al\mbox{.}(2022)]%
        {urooj2022ransomware}
\bibfield{author}{\bibinfo{person}{Umara Urooj}, \bibinfo{person}{Bander Ali~Saleh Al-rimy}, \bibinfo{person}{Anazida Zainal}, \bibinfo{person}{Fuad~A Ghaleb}, {and} \bibinfo{person}{Murad~A Rassam}.} \bibinfo{year}{2022}\natexlab{}.
\newblock \showarticletitle{Ransomware detection using the dynamic analysis and machine learning: A survey and research directions}.
\newblock \bibinfo{journal}{\emph{Applied Sciences}} \bibinfo{volume}{12}, \bibinfo{number}{1} (\bibinfo{year}{2022}), \bibinfo{pages}{172}.
\newblock


\bibitem[Voris et~al\mbox{.}(2019)]%
        {DBLP:journals/compsec/VorisSSHS19}
\bibfield{author}{\bibinfo{person}{Jonathan Voris}, \bibinfo{person}{Yingbo Song}, \bibinfo{person}{Malek~Ben Salem}, \bibinfo{person}{Shlomo Hershkop}, {and} \bibinfo{person}{Salvatore~J. Stolfo}.} \bibinfo{year}{2019}\natexlab{}.
\newblock \showarticletitle{Active authentication using file system decoys and user behavior modeling: results of a large scale study}.
\newblock \bibinfo{journal}{\emph{Comput. Secur.}}  \bibinfo{volume}{87} (\bibinfo{year}{2019}).
\newblock
\urldef\tempurl%
\url{https://doi.org/10.1016/j.cose.2018.07.021}
\showDOI{\tempurl}


\bibitem[Yamany et~al\mbox{.}(2022)]%
        {yamany2022new}
\bibfield{author}{\bibinfo{person}{Bahaa Yamany}, \bibinfo{person}{Mahmoud~Said Elsayed}, \bibinfo{person}{Anca~D Jurcut}, \bibinfo{person}{Nashwa Abdelbaki}, {and} \bibinfo{person}{Marianne~A Azer}.} \bibinfo{year}{2022}\natexlab{}.
\newblock \showarticletitle{A New Scheme for Ransomware Classification and Clustering Using Static Features}.
\newblock \bibinfo{journal}{\emph{Electronics}} \bibinfo{volume}{11}, \bibinfo{number}{20} (\bibinfo{year}{2022}), \bibinfo{pages}{3307}.
\newblock


\bibitem[Zhou et~al\mbox{.}(2023)]%
        {zhou2023limits}
\bibfield{author}{\bibinfo{person}{Chijin Zhou}, \bibinfo{person}{Lihua Guo}, \bibinfo{person}{Yiwei Hou}, \bibinfo{person}{Zhenya Ma}, \bibinfo{person}{Quan Zhang}, \bibinfo{person}{Mingzhe Wang}, \bibinfo{person}{Zhe Liu}, {and} \bibinfo{person}{Yu Jiang}.} \bibinfo{year}{2023}\natexlab{}.
\newblock \showarticletitle{Limits of I/O Based Ransomware Detection: An Imitation Based Attack}. In \bibinfo{booktitle}{\emph{2023 IEEE Symposium on Security and Privacy (SP)}}. IEEE Computer Society, \bibinfo{pages}{2584--2601}.
\newblock


\bibitem[Zhu et~al\mbox{.}(2022)]%
        {DBLP:journals/compsec/ZhuJSWAC22}
\bibfield{author}{\bibinfo{person}{Jinting Zhu}, \bibinfo{person}{Julian Jang{-}Jaccard}, \bibinfo{person}{Amardeep Singh}, \bibinfo{person}{Ian Welch}, \bibinfo{person}{Harith Al{-}Sahaf}, {and} \bibinfo{person}{Seyit Camtepe}.} \bibinfo{year}{2022}\natexlab{}.
\newblock \showarticletitle{A few-shot meta-learning based siamese neural network using entropy features for ransomware classification}.
\newblock \bibinfo{journal}{\emph{Comput. Secur.}}  \bibinfo{volume}{117} (\bibinfo{year}{2022}), \bibinfo{pages}{102691}.
\newblock
\urldef\tempurl%
\url{https://doi.org/10.1016/j.cose.2022.102691}
\showDOI{\tempurl}


\end{thebibliography}

\newpage
\appendix

\section{Sensitivity Analysis}\label{appendix_a}
We conduct sensitivity analysis on two key parameters: the sliding window size for ransom note gene fragments and the similarity matching threshold for ransomware detection. 

\subsection{Sliding Window Size Sensitivity Analysis}
To determine the optimal sliding window size for ransom note gene fragments, we analyzed the system's performance using 158 ransom note samples and 100 benign samples. For each sliding window size, we selected the 300 most frequently occurring combinations as gene fragments. \autoref{tab:sliding_window} presents the results of this analysis.

\begin{table}[htbp]
\centering
\caption{Sliding Window Size Sensitivity Analysis Results}
\def\arraystretch{1.25}
\fontsize{9}{11}\selectfont
\label{tab:sliding_window}
\begin{tabular}{c|c||c|c|c}
\hline
\textbf{Ransomware} & \textbf{Benign} & \textbf{Size} & \textbf{Threshold} & \textbf{Recall} \\
\hline
\multirow{6}{*}{158} & \multirow{6}{*}{100} & 1 & 131 & 6.32\% \\
\cline{3-5}
 &  & 2 & 55 & 34.18\% \\
\cline{3-5}
 &  & 3 & 4 & 84.17\% \\
\cline{3-5}
 &  & 4 & 1 & 81.64\% \\
\cline{3-5}
 &  & 5 & 0 & 68.35\% \\
\cline{3-5}
 &  & 6 & 0 & 56.32\% \\
\hline
\end{tabular}
\end{table}

For each window size, we determined the threshold as the number of gene fragments that resulted in zero false positives in the benign samples. The analysis reveals that the system's performance is highly sensitive to the sliding window size, with the recall rate peaking at 84.17\% for a window size of 3, while maintaining zero false positives.

\subsection{Similarity Matching Threshold Sensitivity Analysis}

\begin{figure}[hbp]
\centering
\includegraphics[width=0.8\linewidth]{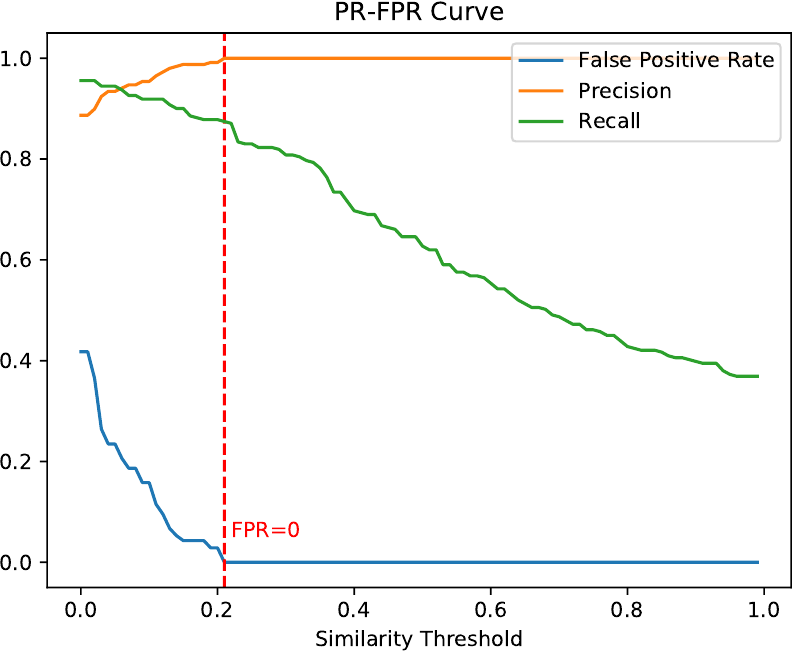}
\caption{Sensitivity analysis of precision, recall, and false alarm rate for the ransom note classifier. The optimal threshold $\tau_{sim} = 0.21$ achieves a 0\% FPR with 100\% precision.}
\label{fig6}
\end{figure}

We also conducted a sensitivity analysis on the similarity matching threshold $\tau_{sim}$ to optimize ransomware detection. This analysis focused on the trade-off between precision, recall, and false alarm rate as the threshold varies.
\autoref{fig6} illustrates how these performance metrics change with different similarity matching thresholds. The analysis reveals that the system's performance is particularly sensitive to the similarity-matching threshold. At $\tau_{sim} = 0.21$, we achieve an optimal balance: a false alarm rate of 0\% while detecting over 87\% of ransomware samples with 100\% precision.

\section{Behavioral Feature Distribution}\label{appendix_b}
\begin{figure}[hbp]
\centering
\includegraphics[width=0.85\linewidth]{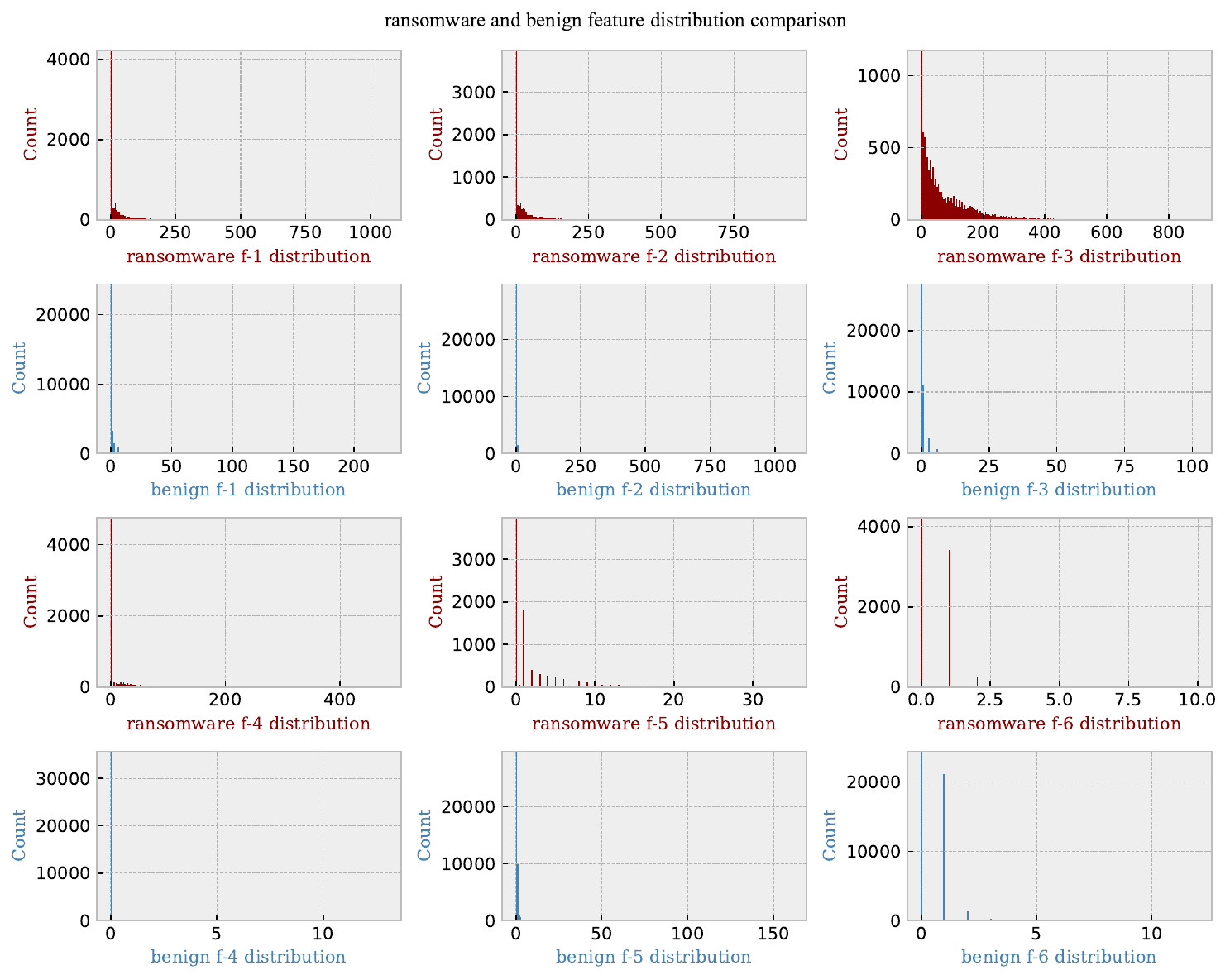}
\caption{Distribution differences for Feature\#1\string~Feature\#6.}
\label{fig8}
\end{figure}

\begin{figure}[t]
\centering
\includegraphics[width=0.85\linewidth]{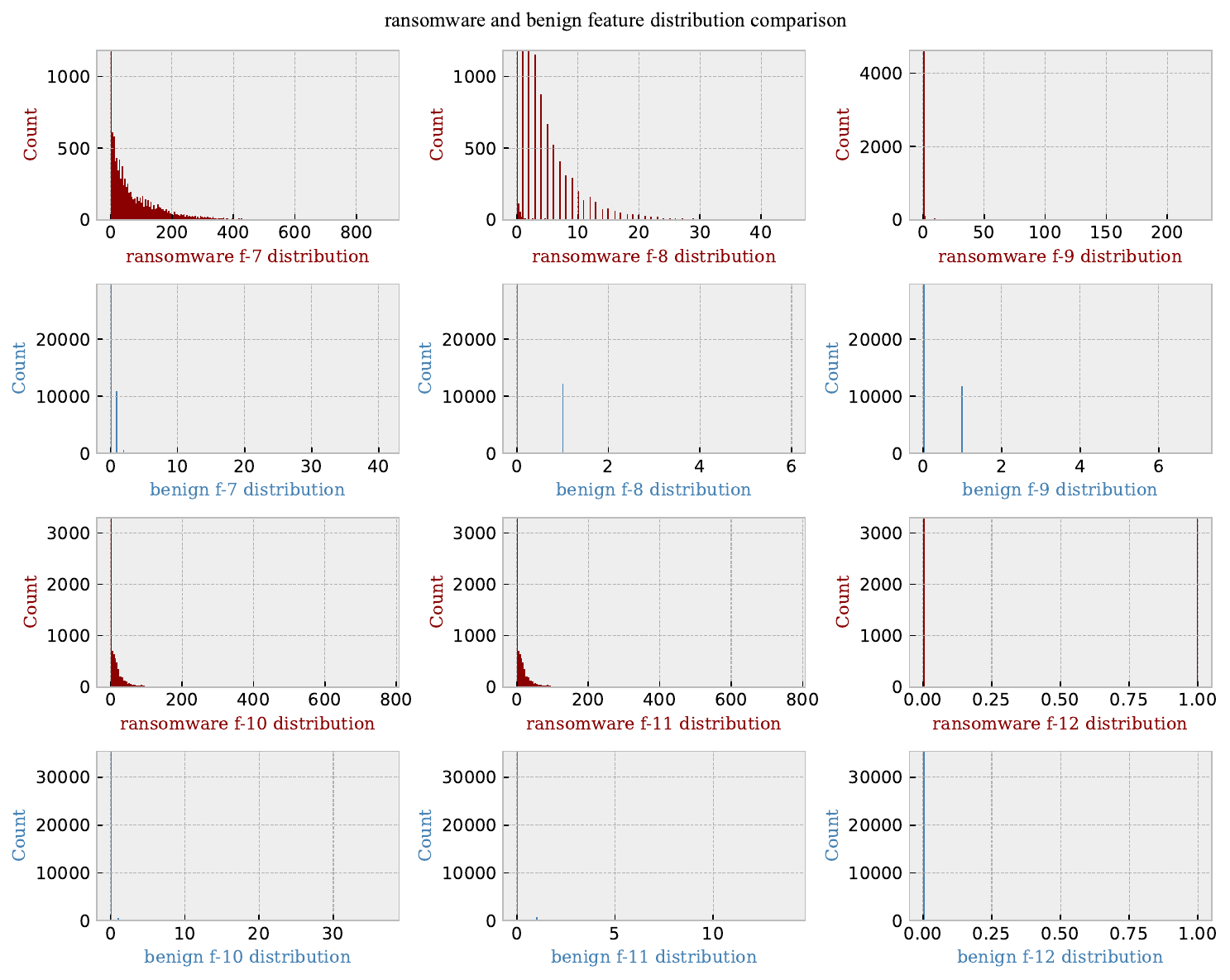}
\caption{Distribution differences for Feature\#7\string~Feature\#12.}
\label{fig9}
\end{figure}

To further explore the behavioral characteristics of the extortion samples, we plot the 12-dimensional feature distribution of benign and ransomware samples in \autoref{fig8} and \autoref{fig9}. The horizontal axis represents the feature values, and the vertical axis represents the amount of benign/ransomware numbers at the corresponding feature. 
The 12-dimensional features correspond to the definition description of \autoref{feature}. 
    
The analysis reveals that the ransomware samples show extremely strong distribution differences from the benign samples in Feature\#3, Feature\#8, and Feature\#11.
Therefore, the feature engineering in this paper can distinguish the benign samples from the ransomware samples well.


\section{Model Selection}\label{appendix_c}

To determine the most effective machine learning model for our ransomware detection system, we conducted a comprehensive evaluation of various algorithms. We used a total of 87,843 behavior data samples, split into 70,274 (80\%) for training and 17,569 (20\%) for validation. Our analysis employed two popular machine learning libraries: scikit-learn and XGBoost. All models were tested using their default parameters to ensure a fair comparison. \autoref{tab:model_comparison} presents the performance metrics for each model tested.

\begin{table}[htbp]
\centering
\caption{Comparison of Machine Learning Models for Ransomware Detection}
\label{tab:model_comparison}
\begin{tabular}{c||c|c|c|c}
\hline
\textbf{Model} & \textbf{Acc} & \textbf{Pre} & \textbf{Rec} & \textbf{F1 Score} \\
\hline
GradientBoosting & 99.23\% & 96.49\% & \textbf{97.80\%} & 97.14\% \\
LogisticRegression & 98.33\% & 90.26\% & 97.30\% & 93.65\% \\
SVM & 97.84\% & 88.80\% & 94.99\% & 91.79\% \\
DecisionTree & 99.30\% & 97.83\% & 97.06\% & 97.44\% \\
XGBoost & \textbf{99.38\%} & \textbf{98.66\%} & 96.84\% & \textbf{97.74\%} \\
RandomForest & 99.36\% & 98.20\% & 97.15\% & 97.67\% \\
ExtraTrees & 99.36\% & 98.16\% & 97.19\% & 97.67\% \\
NaiveBayes & 91.81\% & 78.65\% & 66.98\% & 72.34\% \\
AdaBoosting & 98.22\% & 89.76\% & 96.93\% & 93.21\% \\
\hline
\end{tabular}
\end{table}

Our analysis shows that ensemble methods, particularly tree-based models, demonstrated superior performance across most metrics. Among these, the XGBoost algorithm, an efficient implementation of Gradient Boosting Decision Trees (GBDT), achieved the highest accuracy (99.38\%), precision (98.66\%), and F1 Score (97.74\%). Given its robust and balanced performance in ransomware detection, we selected XGBoost as our primary model for the ransomware detection system.

\section{Detailed Monthly Test Results}\label{appendix_d}
This appendix summarizes the continuous monthly test results from February to June 2023. The results shown in \autoref{tab:appendix-monthly} are obtained from different antivirus software, namely \toolname, Solution A, Solution B, Solution C, and Solution D. The detection results are categorized into two groups: non-process injection categories and process injection categories.
Overall, the statistical results show that \toolname has the highest detection rate for most ransomware families, while Solution A has the highest detection rate for some specific families. Solution B and Solution C have the lowest detection rates for most families. 

\section{Realworld Ransomware Detected}\label{appendix_e}
As statistics listed in \autoref{tab:realworld-all}, \toolname has successfully detected and thwarted 61 real-world ransomware attacks from March 2023 to April 2024. The detected ransomware families include Phobos, Mallox, Lockbit 3.0, and Tellyouthepass. For further analysis, we conducted a manual forensic analysis of 27 ransomware attacks detected from March to June 2023. These cases in \autoref{tab:realworld-forensics} showcase \toolname's remarkable ability to identify and defend against various ransomware families and variants, regardless of whether the attacks were conducted through disguise, vulnerability exploitation, process injection, or other advanced techniques. 
Notably, \toolname successfully prevented 13 attacks leveraging n-day vulnerabilities documented in the CVE database, such as XVE-2023-3377 and QVD-2023-14179, as well as 5 high-risk attacks exploiting zero-day vulnerabilities. This further demonstrates \toolname's outstanding capability in combating unknown and complex ransomware threats.

\begin{table}[hbp]
\centering
\caption{Ransomware attacks detected in each month from March 2023 to April 2024. These ransomware are categorized by their families.}
\label{tab:realworld-all}
\def\arraystretch{1.35}
\fontsize{8.5}{11}\selectfont
\setlength{\tabcolsep}{12pt} 
\begin{tabular}{c|c|c}
\hline
\textbf{Month} & \textbf{Family} & \textbf{Count} \\
\hline
\multirow{2}{*}{March 2023} & Phobos & 1 \\
\cline{2-3}
            & Mallox & 4 \\
\hline
\multirow{2}{*}{May 2023} & Mallox & 5 \\
\cline{2-3}
          & Tellyouthepass & 5 \\
\hline
\multirow{3}{*}{June 2023} & Mallox & 1 \\
\cline{2-3}
           & Tellyouthepass & 10 \\
\cline{2-3}
           & Lockbit3.0 & 1 \\
\hline
\multirow{3}{*}{July 2023} & Mallox & 2 \\
\cline{2-3}
           & Tellyouthepass & 9 \\
\cline{2-3}
           & Lockbit 3.0 & 1 \\
\hline
\multirow{2}{*}{August 2023} & Mallox & 4 \\
\cline{2-3}
             & Tellyouthepass & 3 \\
\hline
\multirow{1}{*}{September 2023} & Mallox & 3 \\
\hline
\multirow{1}{*}{October 2023} & Mallox & 2 \\
\hline
\multirow{2}{*}{November 2023} & Mallox & 1 \\
\cline{2-3}
               & Tellyouthepass & 3 \\
\hline
\multirow{1}{*}{December 2023} & Phobos & 2 \\
\hline
\multirow{2}{*}{January 2024} & Mallox & 1 \\
\cline{2-3}
              & Encrypted & 1 \\
\hline
\multirow{1}{*}{March 2024} & Tellyouthepass & 1 \\
\hline
\multirow{1}{*}{April 2024} & Mallox & 1 \\
\hline
\end{tabular}
\end{table}

\begin{table*}[hbp]
\centering
\caption{Detailed comparison results with four SOTA industrial solutions from February to June 2023.}
\label{tab:appendix-monthly}
\def\arraystretch{1.13}
\fontsize{8}{10}\selectfont
\begin{tabular}{c|c|c|c|cccccc}
\hline
\textbf{Month} & \textbf{Injection Type} & \textbf{Family} & \textbf{Quantity} & \textbf{\toolname} & \textbf{Solution A} & \textbf{Solution B} & \textbf{Solution C} & \textbf{Solution D} \\
\hline
\multirow{6}{*}{February 2023} & \multirow{6}{*}{Non-Process Injection} & BianLian     & 4        & \textbf{4/4 (100.00\%)} & 4/4       & 2/4      & 3/4 &   3/4\\
& & LockbitGreen & 4        & \textbf{4/4 (100.00\%)} & 3/4       & 0/2      & 4/4 &   4/4\\
& & Putin        & 4        &\textbf{4/4 (100.00\%)} & 4/4 & 4/4      & 4/4 &   4/4\\
& & Conti        & 5        & \textbf{4/5 (80.00\%)}  & 4/5       & 0/5      & 5/5 &   4/5\\
& & Lockbit      & 8        & \textbf{8/8 (100.00\%)} & 7/8       & 7/8      & 7/8 &   8/8\\
& & Mimic        & 8        & \textbf{8/8 (100.00\%)} & 8/8       & 7/8      & 8/8 &   8/8\\
\hline
\multirow{5}{*}{March 2023} & \multirow{5}{*}{Non-Process Injection} & Netwalker	& 3   & \textbf{2/3 (66.67\%)}   	& 3/3       & 1/3      & 0/3    & 0/3     \\
& & Ryuk        & 5   & \textbf{5/5 (100.00\%)}    	& 4/5       & 0/5      & 5/5    & 3/5     \\
& & Phobos      & 8   & \textbf{8/8 (100.00\%)}    	& 8/8       & 8/8      & 8/8    & 8/8     \\
& & Chaos       & 14  & \textbf{14/14 (100.00\%)}   & 14/14     & 1/14     & 13/14  & 13/14   \\
& & Stop/Djvu   & 19  & \textbf{19/19 (100.00\%)}   & 18/19     & 19/19    & 19/19 	& 19/19   \\
\hline
\multirow{15}{*}{April 2023} & \multirow{10}{*}{Non-Process Injection} & Blackcat & 1   & \textbf{1/1 (100.00\%)} & 1/1  & 1/1 & 1/1 & 1/1 \\
& & Cylance        & 1   & \textbf{1/1 (100.00\%)} & 1/1  & 0/1 & 1/1 & 1/1 \\
& & DarkPower      & 2   & \textbf{2/2 (100.00\%)} & 2/2  & 0/2 & 1/2 & 2/2 \\
& & Hermes         & 2   & \textbf{2/2 (100.00\%)} & 2/2  & 2/2 & 0/2 & 2/2 \\
& & KadavroVector  & 3   & \textbf{3/3 (100.00\%)} & 3/3  & 1/3 & 3/3 & 2/3 \\
& & MoneyMessage   & 4   & \textbf{4/4 (100.00\%)} & 4/4  & 4/4 & 3/4 & 4/4 \\
& & PayMe100USD    & 2   & \textbf{2/2 (100.00\%)} & 2/2  & 0/2 & 2/2 & 0/2 \\
& & Play           & 1   & \textbf{1/1 (100.00\%)} & 1/1  & 1/1 & 0/1 & 1/1 \\
& & Rcru64         & 8   & \textbf{8/8 (100.00\%)} & 8/8  & 8/8 & 8/8 & 8/8 \\
& & RedEye         & 1   & \textbf{1/1 (100.00\%)} & 1/1  & 1/1 & 1/1 & 1/1 \\
\cline{2-9}
& \multirow{5}{*}{Process Injection} & Ryuk      & 11 & \textbf{11/11 (100.00\%)} & 11/11 & 11/11 & 0/11 & 0/11 \\
& & Waiting   & 2  & \textbf{2/2 (100.00\%)}   & 2/2   & 2/2   & 0/2  & 0/2  \\
& & Magniber  & 6  & \textbf{6/6 (100.00\%)}   & 6/6   & 6/6   & 0/6  & 0/6  \\
& & Netwalker & 10 & \textbf{10/10 (100.00\%)} & 10/10 & 0/10  & 0/10 & 0/10 \\
& & Egregor   & 2  & \textbf{2/2 (100.00\%)}   & 2/2   & 2/2   & 0/2  & 2/2 \\
\hline
\multirow{15}{*}{May 2023} & \multirow{9}{*}{Non-Process Injection} & Akira    & 5 & \textbf{5/5 (100.00\%)} & 5/5 & 5/5 & 5/5 & 5/5 \\
& & Blackbit & 5 & \textbf{5/5 (100.00\%)} & 5/5 & 5/5 & 5/5 & 5/5 \\
& & Conti    & 7 & \textbf{7/7 (100.00\%)} & 7/7 & 7/7 & 7/7 & 7/7 \\
& & Gazprom  & 3 & \textbf{3/3 (100.00\%)} & 3/3 & 3/3 & 0/2 & 3/3 \\
& & Lockbit  & 6 & \textbf{6/6 (100.00\%)} & 6/6 & 6/6 & 6/6 & 5/6 \\
& & Locky    & 3 & \textbf{3/3 (100.00\%)} & 3/3 & 3/3 & 3/3 & 3/3 \\
& & Paradise & 2 & \textbf{2/2 (100.00\%)} & 2/2 & 2/2 & 2/2 & 2/2 \\
& & Play     & 3 & \textbf{3/3 (100.00\%)} & 3/3 & 3/3 & 3/3 & 3/3 \\
& & Rancoz   & 2 & \textbf{2/2 (100.00\%)} & 2/2 & 2/2 & 2/2 & 2/2 \\
\cline{2-9}
& \multirow{6}{*}{Process Injection} & Waiting   & 2 & \textbf{2/2 (100.00\%)} & 2/2 & 2/2 & 0/2 & 0/2 \\
& & Lockbit   & 2 & \textbf{2/2 (100.00\%)} & 2/2 & 0/2 & 2/2 & 1/2 \\
& & Netwalker & 2 & \textbf{2/2 (100.00\%)} & 2/2 & 0/2 & 0/2 & 0/2 \\
& & CrossLock & 2 & \textbf{2/2 (100.00\%)} & 2/2 & 1/2 & 2/2 & 2/2 \\
& & Magniber  & 2 & \textbf{2/2 (100.00\%)} & 1/2 & 2/2 & 0/2 & 0/2 \\
& & Ryuk      & 2 & \textbf{2/2 (100.00\%)} & 1/2 & 1/2 & 0/2 & 1/2 \\
\hline
\multirow{11}{*}{June 2023} & \multirow{5}{*}{Non-Process Injection} & Rhysida     & 2  & \textbf{2/2 (100.00\%)}   & 2/2   & 0/2   & 2/2   & 2/2   \\
& & Darkrace    & 2  & \textbf{2/2 (100.00\%)}  & 1/2   & 0/2   & 2/2   & 2/2   \\
& & WannaCry3.0 & 6  & \textbf{6/6 (100.00\%)}   & 6/6   & 6/6   & 6/6   & 0/6   \\
& & BigHead     & 10 & \textbf{10/10 (100.00\%)} & 10/10 & 10/10 & 10/10 & 3/10  \\
& & BlackBasta  & 30 & \textbf{30/30 (100.00\%)} & 29/30 & 30/30 & 22/30 & 29/30 \\
\cline{2-9}
& \multirow{6}{*}{Process Injection} & Mallox    & 1 & \textbf{1/1 (100.00\%)} & 1/1 & 0/1 & 0/1 & 1/1 \\
& & NoEscape  & 5 & \textbf{5/5 (100.00\%)} & 5/5 & 5/5 & 5/5 & 4/5 \\
& & Phobos    & 2 & \textbf{2/2 (100.00\%)} & 2/2 & 0/2 & 0/2 & 0/2 \\
& & Egregor   & 2 & \textbf{2/2 (100.00\%)} & 2/2 & 2/2 & 0/2 & 2/2 \\
& & Magniber  & 2 & \textbf{2/2 (100.00\%)} & 0/2 & 2/2 & 0/2 & 0/2 \\
& & Netwalker & 2 & \textbf{2/2 (100.00\%)} & 2/2 & 0/2 & 0/2 & 0/2 \\
\hline
\end{tabular}%
\end{table*}

\begin{table*}[hbp]
\centering	
\caption{The detailed manual forensic analysis of real-world ransomware attacks detected from March to June 2023.}
\label{tab:realworld-forensics}
\def\arraystretch{1.35}
\fontsize{9}{12}\selectfont
\setlength{\tabcolsep}{7pt}
\begin{threeparttable}
\begin{tabular}{ccccc}
\hline
\textbf{Date} & \textbf{Industry}         & \textbf{Attack Method\footnote{}}                   & \textbf{Family} & \textbf{Security Advisories} \\ \hline
2023.03.27    & Chemical Industry         & A1                      &  Phobos         &-                     \\
2023.04.11    & Business Services         & A1                      &  Mallox         &-                     \\
2023.05.01    & Retail Industry           & A1                      &  Mallox         &-                     \\
2023.05.02    & Equipment Manufacturing   & A1                      &  Mallox         &-                     \\
2023.05.03    & Construction Engineering  & A1                      &  Mallox         &-                     \\
2023.05.08    & Electronics Manufacturing & A2, A3 &  Tellyouthepass    & XVE-2023-3377      \\
2023.05.10    & Apparel Manufacturing     & A1                      &  Mallox         &-                     \\
2023.05.15    & General Services          & A2,  A3 &  Tellyouthepass    &0-day vulnerability \\
2023.05.15    & Electronics Manufacturing & A2, A3 &  Tellyouthepass    &0-day vulnerability \\
2023.05.15    & Pet Hospital              & A2, A3 &  Tellyouthepass    &0-day vulnerability \\
2023.05.15    & Biomedical Technology     & A2, A3 &  Tellyouthepass    &0-day vulnerability \\
2023.05.15    & Smart Manufacturing       & A2, A3 &  Tellyouthepass    &0-day vulnerability \\
2023.05.22    & Wholesale \& Retail      & A1                      &  Mallox         &-                      \\
2023.05.23    & Biomedical Technology     & A4                         &  Lockbit3.0     &-                       \\
2023.06.09    & Data Bureau               & A2, A3 &  Tellyouthepass    &XVE-2023-3377       \\
2023.06.10    & Investment Management     & A2, A3 &  Tellyouthepass    &XVE-2023-3377       \\
2023.06.18    & Coal Mining               & A3                             &  Mallox         &-                           \\
2023.06.23    & Scientific Research       & A2, A3 &  Tellyouthepass    &QVD-2023-14179      \\
2023.06.23    & General Services          & A2, A3 &  Tellyouthepass    &QVD-2023-14179      \\
2023.06.23    & Smart Transportation      & A2, A3 &  Tellyouthepass    &QVD-2023-14179      \\
2023.06.23    & Equipment Manufacturing   & A2, A3 &  Tellyouthepass    &QVD-2023-14179      \\
2023.06.23    & Home Furnishings          & A2, A3 &  Tellyouthepass    &QVD-2023-14179      \\
2023.06.25    & Chemical Industry         & A2, A3 &  Tellyouthepass    &XVE-2023-3377       \\
2023.06.25    & Construction Engineering  & A2, A3 &  Tellyouthepass    &XVE-2023-3377       \\
2023.06.29    & Equipment Manufacturing   & A2, A3 &  Tellyouthepass    &QVD-2023-14179      \\
2023.06.29    & Internet                  & A2, A3 &  Tellyouthepass    &QVD-2023-14179      \\
2023.06.29    & Scientific Research       & A2, A3 &  Tellyouthepass    &QVD-2023-14179      \\ 
\hline
\end{tabular}
\begin{tablenotes}
  \item[7]{We analyzed four main attack methods. A1: Disguise as normal file; A2: Vulnerability exploitation; A3: Process injection; A4: Add malicious files to the trust zone.} 
\end{tablenotes}
\end{threeparttable}
\end{table*}

\end{document}